\documentclass[aps,11pt,nofootinbib,groupedaddress,preprintnumbers,superscriptaddress]{revtex4}
\usepackage[dvipdfmx]{graphicx}
\usepackage{amssymb,amsfonts,amsmath,bm,color,multirow,cases,empheq,hyperref}
\usepackage{mathrsfs}

\usepackage{enumerate}
\usepackage{ulem}
\usepackage{afterpage}
\hypersetup{%
 setpagesize=false,
 bookmarksnumbered=true,%
 bookmarksopen=true,%
 colorlinks=true,%
 linkcolor=blue,%
 citecolor=red}
\begin{document}

%%%%%%%%%%%%%%%%%%%%%%%%%%%%%%%%%%%%%%%%%%%%
%%%%%%%%%%%%%%%%%%%%%%%%%%%%%%%%%%%%%%%%%%%%
%%%%%%%%%%%%%%%%%%%%%%%%%%%%%%%%%%%%%%%%%%%%

%                                << Title>>

%%%%%%%%%%%%%%%%%%%%%%%%%%%%%%%%%%%%%%%%%%%%
%%%%%%%%%%%%%%%%%%%%%%%%%%%%%%%%%%%%%%%%%%%%
%%%%%%%%%%%%%%%%%%%%%%%%%%%%%%%%%%%%%%%%%%%%

\title{
Stable bound orbits in black lens backgrounds
}

\author{Shinya Tomizawa}
\email{tomizawa@toyota-ti.ac.jp}
\affiliation{Mathematical Physics Laboratory, Toyota Technological Institute, Nagoya 468-8511, Japan}
\author{Takahisa Igata}
\email{igata@post.kek.jp}
\affiliation{KEK Theory Center, Institute of Particle and Nuclear Studies, High Energy Accelerator Research Organization, Tsukuba 305-0801, Japan
}

\preprint{TTI-MATHPHYS-2}
\preprint{KEK-TH-2281, KEK-Cosmo-0269}

%%%%%%%%%%%%%%%%%%%%%%%%%%%%%%%%%%%%%%%%%%%%
%%%%%%%%%%%%%%%%%%%%%%%%%%%%%%%%%%%%%%%%%%%%
%%%%%%%%%%%%%%%%%%%%%%%%%%%%%%%%%%%%%%%%%%%%

%                                << abstract>>

%%%%%%%%%%%%%%%%%%%%%%%%%%%%%%%%%%%%%%%%%%%%
%%%%%%%%%%%%%%%%%%%%%%%%%%%%%%%%%%%%%%%%%%%%
%%%%%%%%%%%%%%%%%%%%%%%%%%%%%%%%%%%%%%%%%%%%

\begin{abstract} 
 In contrast to five-dimensional Schwarzschild-Tangherlini and Myers-Perry backgrounds, we show that there are stable bound orbits of massive/massless particles in five-dimensional black lens backgrounds, in particular, the supersymmetric black lens with $L(2,1)$ and $L(3,1)$ topologies. We also show that in the zero-energy limit of massless particles, there exist stable circular orbits on the evanescent ergosurfaces.
\end{abstract}

\pacs{04.50.+h  04.70.Bw}
\date{\today}
\maketitle

%%%%%%%%%%%%%%%%%%%%%%%%%%%%%%%%%%%%%%%%%%%%
%%%%%%%%%%%%%%%%%%%%%%%%%%%%%%%%%%%%%%%%%%%%
%%%%%%%%%%%%%%%%%%%%%%%%%%%%%%%%%%%%%%%%%%%%

%                                << Introduction>>

%%%%%%%%%%%%%%%%%%%%%%%%%%%%%%%%%%%%%%%%%%%%
%%%%%%%%%%%%%%%%%%%%%%%%%%%%%%%%%%%%%%%%%%%%
%%%%%%%%%%%%%%%%%%%%%%%%%%%%%%%%%%%%%%%%%%%%
\section{Introduction}

A stable bound particle orbit is an orbit where a particle keeps moving in a bounded spatial region without reaching infinity or singularities even if small perturbations are applied. 
Such an orbit often appears in astrophysical phenomena because the stability results in a relatively long duration. 
The essence of the mechanism can be seen from the dynamics of a massive particle in the Schwarzschild spacetime. 
A massive particle moving on a stable bound orbit is localized near a radial potential well, which is made by a balance between the gravitational potential $-Mm/r$ and the centrifugal potential $l^2/(2m^2r^2)$, where $M$ is the black hole mass, $m$ and $l$ are the mass and angular momentum of a particle, and $r$ is the circumferential radius.
The relativistic correction effect $-Ml^2/(m^2r^3)$ becomes dominated near the event horizon and causes proper relativistic phenomena such as perihelion shift and the innermost stable circular orbits. 
These can explain or predict events in the vicinity of black holes and other celestial objects, such as stellar orbital motion and accretion disks.
Similarly, it was shown that stable bound orbits exist in the Kerr black hole spacetime~\cite{Wilkins:1972rs}.

For photons moving in the Schwarzschild spacetime and Kerr spacetime, it is very well-known that there exist unstable circular orbits but not stable ones. 
However, in the Kerr-Newman spacetime with relatively large electric charge, 
stable photon orbits exist on the horizon~\cite{Khoo:2016xqv}, and 
in the 4D Majumdar-Papapetrou spacetimes with two black holes, 
they appear even outside the horizon~\cite{Dolan:2016bxj, Nakashi:2019mvs, Nakashi:2019tbz}.
The existence of such stable bound photon orbits has a physically significant meaning
because it implies instability of the background spacetime in the following sense: 
If they exist, many massless particles (not only photons but also gravitons, etc.) 
are stably trapped on the orbits and accumulate more and more in the finite region of the spacetime. 
This will cause so large backreaction to
 the background geometry that it will eventually break the background. 
On the other hand, from the wave perspective, linear waves localize in the vicinity of the trapping null geodesics resulting in a long timescale for the decay~\cite{Keir:2014oka}.
This phenomenon suggests the existence of nonlinear instabilities of the background spacetime~\cite{Cardoso:2014sna}.

\medskip
It is now evident that even within the framework of vacuum Einstein gravity, there is a much richer variety of black hole solutions in higher dimensions. 
For instance, an asymptotically flat, stationary and biaxisymmetric five-dimensional black holes can have 
three types of topologies of the event horizon, a sphere $S^3$, a ring $S^1\times S^2$ and lens spaces $L(p,q)$ \cite{Cai:2001su,Galloway:2005mf,Hollands:2007aj,Hollands:2010qy}. 
The corresponding solutions are, respectively, called a black hole, a black ring~\cite{Emparan:2001wn,Pomeransky:2006bd,Elvang:2004rt} and a black lens~\cite{Kunduri:2014kja,Tomizawa:2016kjh}. 
For the Schwarzschild-Tangherlini solution and Myers-Perry solution in five dimensions, it was shown that there are no stable circular orbits in equatorial planes~\cite{Tangherlini:1963bw,Frolov:2003en,Page:2006ka,Frolov:2006pe,Cardoso:2008bp}.
For the black ring solutions, it was shown in contrast to black holes that there exist stable bound orbits~\cite{Igata:2010ye,Igata:2010cd,Igata:2013be}.
For the black lens solutions, in our previous work~\cite{Tomizawa:2019egx}, we numerically show examples of stable bound orbits of particles around the supersymmetric black lens with the horizon topology of $L(2,1)$ in the five-dimensional minimal supergravity.

\medskip
In this paper, focusing on the supersymmetric black lenses with $L(2,1)$ and $L(3,1)$ topologies in the five-dimensional minimal supergravity, we give more detail about the existence of the stable bound orbits for the massive and massless particles which move around the horizon than in the previous paper. 
In our analysis, we consider such motion of particles as a two-dimensional potential problem, 
where a problem of whether a stable bound orbit exists for massive or massless particles is reduced to a simple problem of whether the two-dimensional effective potential has a negative or zero local minimum. 
 We also discuss if there are stable bound orbits at infinity because for the five-dimensional Majumdar-Papapetrou solution with two black holes~\cite{Igata:2020vlx}, they can exist at infinity when the separation of two black holes is large enough.

\medskip
The rest of the paper is composed as follows: 
In the following Sec.~\ref{sec:black lens}, we briefly review the supersymmetric black lens solution in the five-dimensional minimal supergravity. 
In Sec.~\ref{sec:formalism}, we provide our formalism to show the existence of stable bound orbits. 
In Sec.~\ref{sec:SBO}, we show that there are stable bound orbits for supersymmetric black lenses with $L(2,1)$ and $L(3,1)$ topologies. 
In Sec.~\ref{sec:summary}, we summarize our results and discuss possible generalizations of our analysis to other black hole backgrounds.

%%%%%%%%%%%%%%%%%%%%%%%%%%%%%%%%%%%%%%%%%%%%
%%%%%%%%%%%%%%%%%%%%%%%%%%%%%%%%%%%%%%%%%%%%
%%%%%%%%%%%%%%%%%%%%%%%%%%%%%%%%%%%%%%%%%%%%

%                                << Black lens solutions>>

%%%%%%%%%%%%%%%%%%%%%%%%%%%%%%%%%%%%%%%%%%%%
%%%%%%%%%%%%%%%%%%%%%%%%%%%%%%%%%%%%%%%%%%%%
%%%%%%%%%%%%%%%%%%%%%%%%%%%%%%%%%%%%%%%%%%%%
\section{Black lenses}\label{sec:black lens}

\subsection{Solutions}

In the five-dimensional minimal supergravity, the local metric and gauge potential $1$-form of the supersymmetric black lens solutions take the form~\cite{Kunduri:2014kja,Tomizawa:2016kjh}
\begin{eqnarray}
\label{metric}
ds^2&=&-f^2(dt+\omega)^2+f^{-1}ds_{M}^2,\\
A&=&\frac{\sqrt 3}{2} \left[f(d t+\omega)-\frac KH(d \psi+\chi)-\xi \right]\,,
\end{eqnarray}
where $ds^2_M$ is the Gibbons-Hawking metric, which is given by
\begin{eqnarray}
ds^2_M&=&H^{-1}(d\psi+\chi)^2+H(dx^2+dy^2+dz^2), \\
\chi&=&\sum_{i=1}^nh_i\frac{z-z_i}{r_i}\frac{xdy-ydx}{x^2+y^2},\\
H&=&\sum_{i=1}^n\frac{h_i}{r_i}:=\frac{n}{r_1}-\sum_{i= 2}^n\frac{1}{r_i}, \label{Hdef}
\end{eqnarray}
where $r_i:=|{\bm r}-{\bm r_i}|=\sqrt{x^2+y^2+(z-z_i)^2}$, ${\bm r}:=(x,y,z)$ and ${\bm r}_i:=(0,0,z_i)$. For the constants $z_i$, we assume $z_1=0<z_2<\cdots<z_n$ and $H$ is a harmonic function with point sources at ${\bm r}={\bm r}_i$ on ${\mathbb E}^3$. 
The vectors $\partial/\partial t$, $\partial/\partial \psi$ and $\partial/\partial \phi:=x\partial/\partial y-y\partial/\partial x$ are Killing vectors, where $\partial/\partial\phi$ is the coordinate basis in the standard polar coordinates $(x=r\sin\theta\cos\phi,y=r\sin\theta\sin\phi,z=r\cos\theta)$. 
The other functions and one forms are written as
\begin{eqnarray}
f^{-1}&=&H^{-1}K^2+L,\\
\omega&=&\omega_\psi(d\psi+\chi)+\hat \omega,\\
\omega_\psi&=&H^{-2}K^3+\frac{3}{2} H^{-1}KL+M, \\
\hat \omega&=&\sum_{j=2}^n\left[\frac{n}{2}k_j^3+\frac{3}{2}(k_1k_j^2-k_jl_1)\right]\frac{r-z_j\cos\theta}{z_{j}r_j}d\phi\notag\\
&&+\sum_{i,j=2(i\not=j)}^n\left(-\frac{1}{2}k_j^3+\frac{3}{2}k_i k_j^2 \right)\frac{r^2-(z_i+z_j)r\cos\theta+z_iz_j}{z_{ji}r_ir_j}d\phi \notag \\
&&-\frac{3}{2}\sum_{i=1}^n\left(-\sum_{j=1}^nk_j h_i+k_i\right)\frac{z-z_i}{r_i}d \phi\notag\\
&&-\sum_{j=2}^n\frac{nk_j^3+3(k_1k_j^2-k_jl_1)}{2z_{j}}d\phi-\sum_{i,j=2(i\not =j)}^n\frac{h_ik_j^3+3k_ik_j^2}{2z_{ji}}d\phi,\\
\xi&=-&\sum_{i=1}^nk_i\frac{z-z_i}{r_i}d \phi,
\end{eqnarray}
where $z_{ji}:=z_j-z_i$ and 
\begin{eqnarray}
K&=&\sum_{i=1}^n\frac{k_i}{r_i},\\
L&=&
1+\frac{l_1}{r_1}+\sum_{i=2}^n\frac{k_i^2}{r_i},\\
M&=&
-\frac{3}{2}\sum_{i=1}^nk_i+\sum_{i= 2}^n\frac{k_i^3}{2r_i}.
\end{eqnarray}

\medskip
%%%%%%%%%%%%%%%%%%%%%%%%%%%%%%%%%%%%%%%%%%%%
%%%%%%%%%%%%%%%%%%%%%%%%%%%%%%%%%%%%%%%%%%%%
%%%%%%%%%%%%%%%%%%%%%%%%%%%%%%%%%%%%%%%%%%%%

%                                << Boundary conditions}>>

%%%%%%%%%%%%%%%%%%%%%%%%%%%%%%%%%%%%%%%%%%%%
%%%%%%%%%%%%%%%%%%%%%%%%%%%%%%%%%%%%%%%%%%%%
%%%%%%%%%%%%%%%%%%%%%%%%%%%%%%%%%%%%%%%%%%%%

As discussed in Ref.~\cite{Tomizawa:2016kjh}, from the requirements of regularity at ${\bm r}={\bm r}_i\ (i=2,\ldots,n)$ and the absence of closed timelike curves around the horizon ${\bm r}={\bm r}_1$ and the $(n-1)$ points ${\bm r}={\bm r}_i\ (i=2,\ldots,n)$, the parameters $(k_{i\ge 1},l_1,z_{i\ge 2})$ must satisfy 
\begin{eqnarray}
1+\frac{1}{z_{i}}(l_1-2k_ik_1-nk_i^2)+\sum_{ j=2(j\not=i)}^n\frac{1}{|z_{ji}|}(k_j-k_i)^2 <0,\label{eq:c1}\\
-\frac{3}{2}\sum_{j=1}^nk_j-\frac{3}{2}k_i+\frac{nk_i^3+3k_1k_i^2-3l_1k_i}{2z_{i}}+\sum_{j=2(j\not =i)}^n\frac{(k_j-k_i)^3}{2|z_{ji}|}=0
\label{eq:c2}
\end{eqnarray}
for $i=2,\ldots,n$ and the inequalities 
\begin{eqnarray}
k_1^2+nl_1> 0,\quad 
l_1^2(3k_1^2+4nl_1)> 0.\label{eq:R1R2ineq}
\end{eqnarray}
It is shown in Ref.~\cite{Tomizawa:2016kjh} that when the parameters simultaneously satisfy these conditions, the point ${\bm r}={\bm r}_1(=0)$ denotes a null degenerate horizon whose spatial cross section is the lens space $L(n,1)$, whereas the points ${\bm r}={\bm r_i}\ (i=2,\ldots,n)$ give regular points, which correspond to the merely coordinate singularities like the origin of the Minkowski spacetime in the Gibbons-Hawking coordinates. 

\subsection{Evanescent ergosurface}

The supersymmetric black lens admits the presence of evanescent ergosurfaces~\cite{Kunduri:2014kja,Tomizawa:2016kjh}, which are defined as timelike hypersurfaces such that a stationary Killing vector field becomes null there and timelike everywhere except there. 
Reference~\cite{Eperon:2016cdd} proved that on such surfaces, massless particles with zero energy ($E=0$) relative to infinity move along stable trapped null geodesics.
They exist at $f=0$, which corresponds to 
\begin{eqnarray}
H=\sum_{i=1}^n\frac{h_i}{r_i}=0.
\end{eqnarray}
For $n=2$, they cross the points $z=2z_2/3$ and $z=2z_2 $ on the $z$ axis.
For $n=3$, they cross the points $z$ satisfying 
\begin{eqnarray}
F(z):=3|z-z_2||z-z_3|-|z||z-z_3|-|z||z-z_2|=0
\end{eqnarray}
on the $z$ axis. It turns out from simple computations that $F(z)=0$ has only a single root on $I_+$ and $I_1$, two roots on $I_2$ and no root on $I_-$.

\subsection{New coordinates}
In the work of the geodesic motion of massive/massless particles around the black lenses, it is more convenient to use the coordinates $(\eta,\xi,\phi_1,\phi_2)$ defined by
\begin{eqnarray}
&&\eta=2\sqrt{r}\cos\frac{\theta}{2},\quad \xi=2\sqrt{r}\sin\frac{\theta}{2},\\
&&\phi_1=\frac{\psi+\phi}{2},\quad \phi_2=\frac{\psi-\phi}{2},
\end{eqnarray}
than the Gibbons-Hawking coordinates $(r,\theta,\phi,\psi)$, 
where $(\phi_1,\phi_2)$ are the coordinates with $2\pi$ periodicity. 
It should be noted that the points ${\bm r}_i=(0,0,z_i)\ (i=1,2,\ldots,n)$ on the $z$ axis correspond to $(\eta,\xi)=(\eta_i,0)\ (i=1,2,\ldots,n)$ on the $\eta$ axis in the new coordinates, where $\eta_i:=2\sqrt{z_i}$.

\subsection{$n=2$ case}
The case $n=2$ coincides with a black lens solution with the horizon topology of $L(2,1)$ in Ref.~\cite{Kunduri:2014kja}. Equation~(\ref{eq:c2}) is simply written as
\begin{eqnarray}
z_2=\frac{k_2 \left(3 k_1 k_2+2 k_2^2-3 l_1\right)}{3 (k_1+2 k_2)}(>0), \label{eq:z2ineq_n=2}
\end{eqnarray}
and the inequalities (\ref{eq:c1}) and (\ref{eq:R1R2ineq}) are, respectively, 
\begin{eqnarray}
 1 + \frac{l_1 - 2 k_2k_1 - 2k_2^2}{z_2}<0, \label{eq:c2ineq_n=2}
\end{eqnarray}
\begin{eqnarray}
l_1^2(3 k_1^2 + 8l_1)>0.\label{eq:R1ineq_n=2}
\end{eqnarray}
The blue-shaded regions ${\cal D}$ in Fig.~\ref{fig:PR2} show the parameter region where the inequalities (\ref{eq:z2ineq_n=2}), (\ref{eq:c2ineq_n=2}) and (\ref{eq:R1ineq_n=2}) are simultaneously satisfied for $l_1=1$ under Eq.~(\ref{eq:z2ineq_n=2}). 
In particular, we consider the case of $k_1=0$, which simplifies the conditions (\ref{eq:z2ineq_n=2})-(\ref{eq:R1ineq_n=2}) as
\begin{eqnarray}
&&z_2=\frac{2k_2^2-3l_1}{6}>0,\\
 &&1+\frac{l_1-2k_2^2}{z_2}<0,\\
&&l_1>0.
\end{eqnarray}

\subsection{$n=3$ case}
\begin{figure}[h]
 \begin{tabular}{cc}
 \begin{minipage}[t]{0.5\hsize}
\includegraphics[width=6cm]{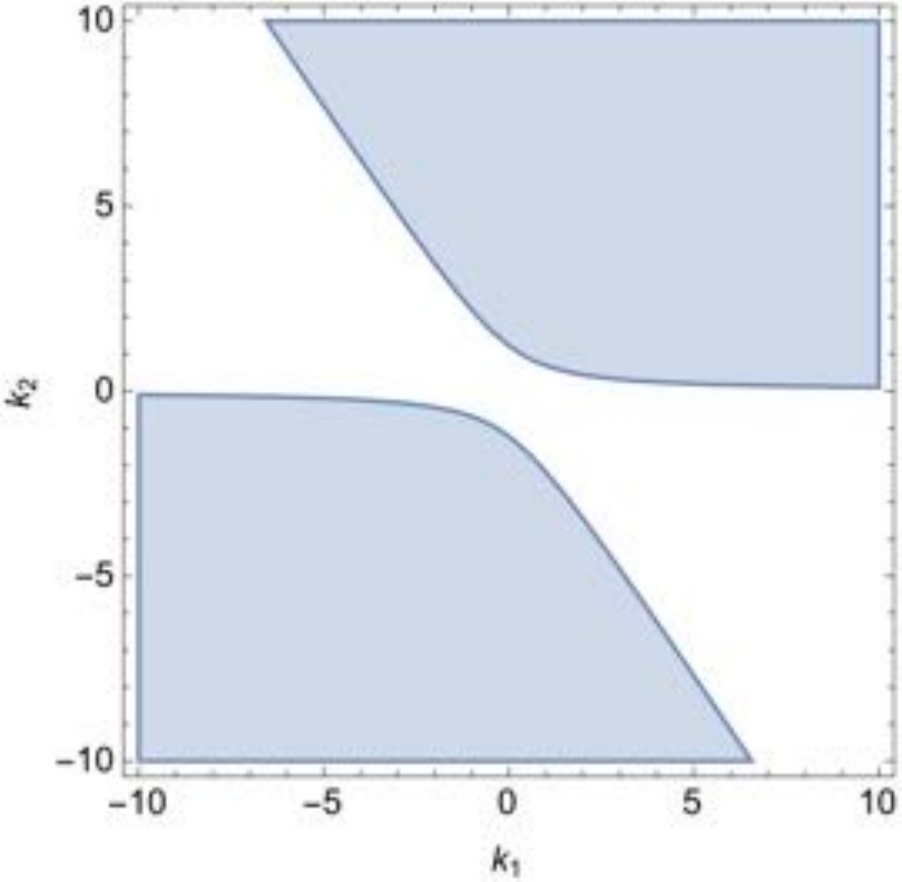}
\caption{Parameter regions ${\cal D}$ for the black lens\\ with $L(2,1)$ topology.}
\label{fig:PR2}
 \end{minipage} &
 
 \begin{minipage}[t]{0.5\hsize}
\includegraphics[width=6cm]{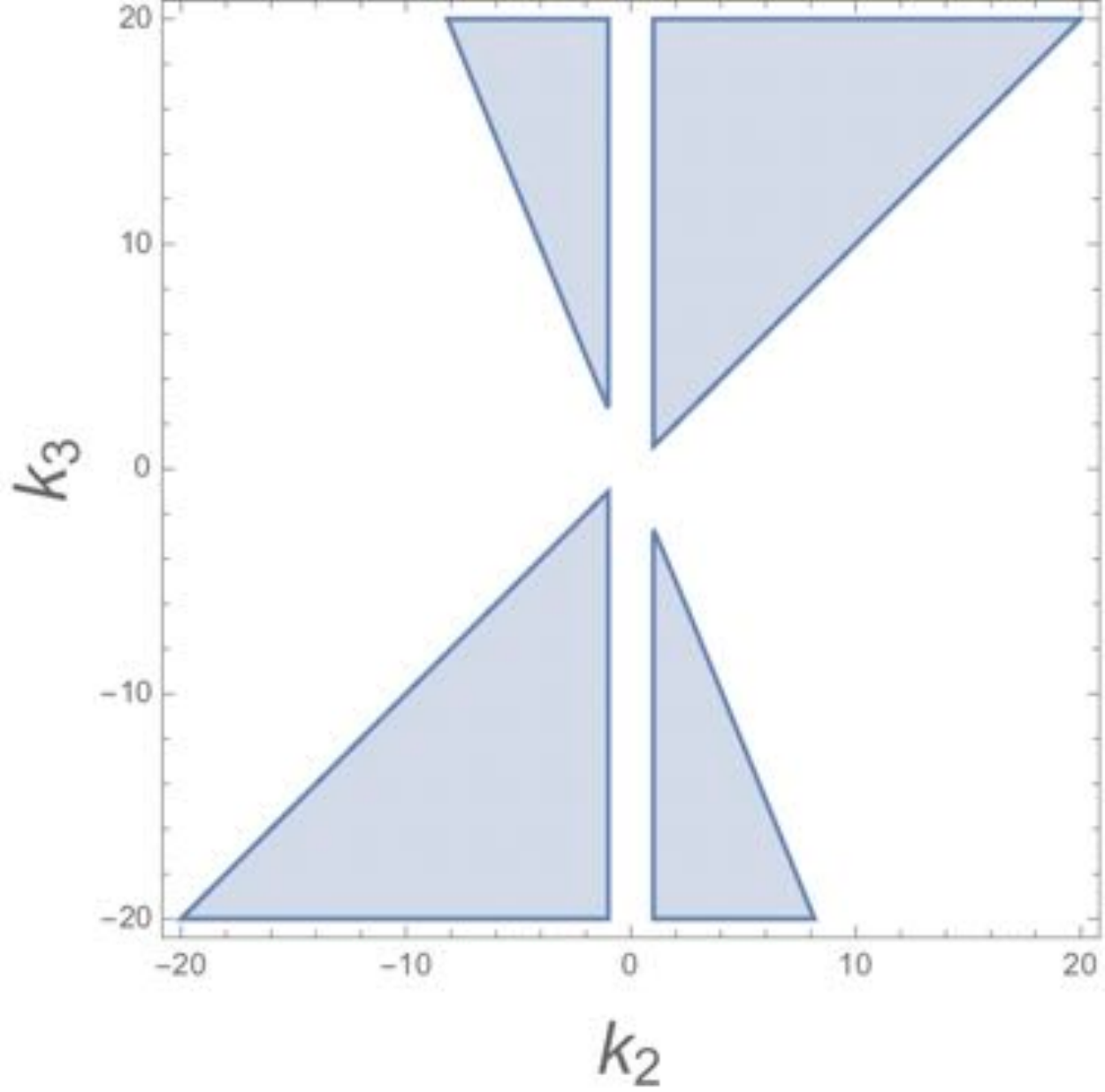}
\caption{Parameter regions ${\cal D}$ for the black lens\\ with $L(3,1)$ topology ($k_1=0$).}
\label{fig:PR3}
 \end{minipage}

\end{tabular}
\end{figure}

The case $n=3$ describes a black lens with the horizon topology of $L(3,1)$, which has $4$ independent parameters $(k_1,k_2,k_3,l_1)$.
In this paper, for simplicity, we consider only the case of $k_1=0$, in which case the conditions~(\ref{eq:c2}) are written as 
\begin{eqnarray}
&&-\frac{3}{2}(2k_2+k_3)+\frac{3(k_2^3-l_1k_2)}{2z_2}+\frac{(k_3-k_2)^3}{2z_{32}}=0, \label{eq:c22}\\
&&-\frac{3}{2}(k_2+2k_3)+\frac{3(k_3^3-l_1k_3)}{2z_3}+\frac{(k_2-k_3)^3}{2z_{32}}=0,\label{eq:c23}
\end{eqnarray}
and the inequalities~(\ref{eq:c1}) and~(\ref{eq:R1R2ineq}) are reduced to, respectively, 
\begin{eqnarray}
&&1+\frac{l_1-3k_2^2}{z_2}+\frac{(k_3-k_2)^2}{z_{32}}<0,\label{eq:c12}\\
&&1+\frac{l_1-3k_3^2}{z_3}+\frac{(k_3-k_2)^2}{z_{32}}<0, \label{eq:c13}\\
&&l_1>0.\label{eq:ctc1}
\end{eqnarray}
From Eqs.~(\ref{eq:c22}) and (\ref{eq:c23}), $z_2$ and $z_3$ can be written as the functions of $k_2$, $k_3$ and $l_1$. When we normalize $l_1=1$ from (\ref{eq:ctc1}), the inequalities~(\ref{eq:c12}) and (\ref{eq:c13}) can be denoted by a certain region in a $(k_2,k_3)$-plane. This parameter region ${\cal D}$ which gives a black lens with $L(3,1)$ topology is drawn as the blue-colored region in Fig.~\ref{fig:PR3}.

%%%%%%%%%%%%%%%%%%%%%%%%%%%%%%%%%%%%%%%%%%%%
%%%%%%%%%%%%%%%%%%%%%%%%%%%%%%%%%%%%%%%%%%%%
%%%%%%%%%%%%%%%%%%%%%%%%%%%%%%%%%%%%%%%%%%%%

%                                << Black Ring solutions>>

%%%%%%%%%%%%%%%%%%%%%%%%%%%%%%%%%%%%%%%%%%%%
%%%%%%%%%%%%%%%%%%%%%%%%%%%%%%%%%%%%%%%%%%%%
%%%%%%%%%%%%%%%%%%%%%%%%%%%%%%%%%%%%%%%%%%%%

%%%%%%%%%%%%%%%%%%%%%%%%%%%%%%%%%%%%%%%%%%%%
%%%%%%%%%%%%%%%%%%%%%%%%%%%%%%%%%%%%%%%%%%%%
%%%%%%%%%%%%%%%%%%%%%%%%%%%%%%%%%%%%%%%%%%%%

%                                << Potential on the $z$ axis>>

%%%%%%%%%%%%%%%%%%%%%%%%%%%%%%%%%%%%%%%%%%%%
%%%%%%%%%%%%%%%%%%%%%%%%%%%%%%%%%%%%%%%%%%%%
%%%%%%%%%%%%%%%%%%%%%%%%%%%%%%%%%%%%%%%%%%%%

\section{Our formalism} \label{sec:formalism}
Our method to find stable bound orbits is based on the previous work~\cite{Tomizawa:2019egx}, where we considered the geodesic motion of particles around the black lenses as a two-dimensional potential problem. 
Now, we give the brief review as follows. 
The Hamiltonian of a free particle with the mass $m$ is written as
\begin{eqnarray}
\mathcal{H}=g^{\mu\nu}p_\mu p_\nu+m^2, \label{eq:Hamiltonian}
\end{eqnarray}
where $p_\mu$ is the momentum. 
From the independence of ${\mathcal H}$ on the coordinates $(t,\phi_1,\phi_2)$, 
the momenta $(p_t,p_{\phi_1},p_{\phi_2})$ are constants of motion, and we denote them as
 $(p_t,p_{\phi_1},p_{\phi_2})=(-E,L_{\phi_1},L_{\phi_2})$. 
Then, the Hamiltonian can be written in terms of these constants as
\begin{eqnarray}
\mathcal{H}=\frac{4f}{H(\eta^2+\xi^2)}(p_\eta^2+p_\xi^2)+E^2\left(U+\frac{m^2}{E^2}\right).\label{eq:Hamiltonian2}
\end{eqnarray}
The function $U$ is the effective potential defined by
\begin{eqnarray}
U&=&g^{tt}+g^{\phi_1\phi_1}l_{\phi_1}^2+g^{\phi_2\phi_2}l_{\phi_2}^2-2g^{t\phi_1} l_{\phi_1}-2g^{t\phi_2} l_{\phi_2}+2g^{\phi_1\phi_2}l_{\phi_1} l_{\phi_2}\\
&=&\frac{1}{4(K^2+HL)}\biggl[ -3K^2L^2+8K^3M+12HKLM-4HL^3+4H^2M^2 \notag\\
&&+ (4K^3+6HKL+4H^2M)(l_{\phi_1}+l_{\phi_2})+H^2(l_{\phi_1}+l_{\phi_2})^2 \biggr]\notag\\
&&+\frac{[-2\hat\omega_\phi +(l_{\phi_1}+l_{\phi_2})\chi_\phi+(l_{\phi_2}-l_{\phi_1})]^2}{(K^2+HL)\eta^2\xi^2},
\end{eqnarray}
where we have normalized two angular momenta by the energy as
 $l_{\phi_1}:=L_{\phi_1}/E$ and $l_{\phi_2}:=L_{\phi_2}/E$. 
Thus we can consider that the massive and massless particles move on the two-dimensional space $(\eta,\xi)$ while satisfying the Hamiltonian constraint ${\mathcal H}=0$. 
When we consider that the particles move in the two-dimensional potential $U$, the allowed regions of the motions for massive and massless particles are restricted to $U\le -m^2/E^2$ and $U\le0$, respectively. 
From the determinant and trace of the Hesse matrix $(H_{ij}):=(U_{,i,j})\ (i,j=\eta,\xi)$ of $U$, 
we can discuss the existence of a local minimum. 
If $\mathrm{tr\ }(H_{ij})>0$ and $\mathrm{det\ }(H_{ij})>0$ at a stationary point $U_{,i}=0$, $U$ has a local minimum at the point.

\medskip

For simplicity, we focus on the shape of the potential $U$ on the $z$ axis (i.e., $\theta=0,\pi$) of ${\mathbb E}^3$ in the Gibbons-Hawking space, which corresponds to $\eta=0$ and $\xi=0$ in the coordinates $(\eta,\xi)$. 
The $z$ axis is composed of the $n+1$ intervals: 
$I_-=\{(\eta,\xi)|\eta=0,\xi>0\}$, $I_i=\{(\eta,\xi)|\eta_i<\eta<\eta_{i+1},\xi=0\}\ (i=1,2,\ldots, n-1)$ and $I_+=\{(\eta,\xi)|\eta>\eta_n,\xi=0\}$. 
On $I_+$ and $I_-$, only the particles with the angular momenta of $l_{\phi_2}=0$ and $l_{\phi_1}=0$, respectively, are allowed to stay there, whereas on $I_i\ (i=1,\ldots n-1)$ only the particles with the angular momenta of the special ratio of $l_{\phi_1}/l_{\phi_2}=-(n-i+1)/(n-i)$ are allowed to stay there because $I_i$ corresponds to the fixed points of the Killing isometry $v:=(n-i)\partial/\partial \phi_1+(n-i+1)\partial/\partial\phi_2$, and hence only the particle with a zero angular momentum of $J:=p_\mu v^\mu=(n-i)L_{\phi_1}+(n-i+1)L_{\phi_2}=0$ can stay on the axis $I_i$ and otherwise the potential diverges on $I_i$.

\medskip
In the above formalism, we must remove the zero-energy limit $E\to 0$ because $l_{\phi_1}$ and $l_{\phi_2}$ are divided by $E$. For this limit, it is better to use the potential $U'$ defined by
\begin{eqnarray}
U'=g^{\phi_1\phi_1}L_{\phi_1}^2+g^{\phi_2\phi_2}L_{\phi_2}^2+2g^{\phi_1\phi_2}L_{\phi_1}L_{\phi_2}
\end{eqnarray}
instead of the potential $U$, though in the previous work~\cite{Tomizawa:2019egx}, we considered the zero-energy as the limit $\l_{\phi_1},\l_{\phi_2}\to \pm\infty$ of $U$. 
The potential $U'$ takes a simple form of
\begin{eqnarray}
U'&=&\frac{H^2(L_{\phi_1}+L_{\phi_2})^2}{4(K^2+HL)}+\frac{[(L_{\phi_2}-L_{\phi_1})+ (L_{\phi_2}+L_{\phi_1})\chi_\phi ]^2}{(K^2+HL)\eta^2\xi^2}.\label{eq:up}
\end{eqnarray}
For this zero energy limit, massive particles are not allowed to move because $U'$ is nonnegative, whereas massless particles can move on $U'=0$, which corresponds to the intersection of two curves in the $(\eta,\xi)$ plane, $H=0$ and $G:=(L_{\phi_2}-L_{\phi_1})+ (L_{\phi_2}+L_{\phi_1})\chi_\phi =0$, because the first and second terms in Eq.~(\ref{eq:up}) are nonnegative. 
 It turns out hence that massless particles with zero-energy always move on the evanescent surfaces, which correspond to $H=0$. 
 Moreover, $U'=0$ corresponds to a stationary point because at the point $\partial_iU'=0\ (i=\eta,\xi)$ also holds. 
 
It can be shown that $U'$ has a local minimum at such a stationary point.
To this end, let us confirm both the determinant and trace of the Hesse matrix $H:=(\nabla_i\nabla_j U')\ (i,j=\eta,\xi)$ are positive at the stationary point, where $\nabla_i$ are the covariant derivatives associated with the two-dimensional conformally flat metric $g_{ij}$ in Eqs.~(\ref{eq:Hamiltonian}) and (\ref{eq:Hamiltonian2}). 
At just a stationary point such that $\nabla_iU'=\partial_iU'=0$, $\nabla_i\nabla_j U=\partial_i\partial_j U'$ can be shown because
\begin{eqnarray}
\nabla_i\nabla_jU'=\partial_i\partial_jU'-\Omega^{-1}(\partial_iU'\partial_j\Omega+\partial_jU'\partial_i\Omega-\delta_{ij} \partial_kU'\partial^k\Omega),
\end{eqnarray}
where $\Omega^2:=H(\eta^2+\xi^2)/4f$, so that 
 it is enough to compute the determinant and trace of $\partial_i\partial_j U$, which are written, respectively, as
\begin{eqnarray}
{\rm det}(\partial_i\partial_j U')\big|_{H=G=0}&=&U'_{\eta\eta}U'_{\xi\xi}-U_{\eta\xi}^{\prime2}\big|_{H=G=0}\notag\\
&=&\frac{(L_{\phi_1}+L_{\phi_2})^2(H_{,\xi}G_{,\eta}-H_{,\eta}G_{,\xi})^2}{K^4\eta^2\xi^2}\notag\\
&=&\frac{(L_{\phi_1}+L_{\phi_2})^4(H_{,\xi}^2+H_{,\eta}^2)^2}{4K^4}>0,\\
{\rm Tr}(\partial_i\partial_j U')\big|_{H=G=0}&=&U'_{\eta\eta}+U'_{\xi\xi}\big|_{H=G=0}\notag\\
                     &=&\frac{4(G_{,\eta}^2+G_{,\xi}^2)+(L_{\phi_1}+L_{\phi_2})^2(H_{,\eta}^2+H_{,\xi}^2)\eta^2\xi^2}{2K^2\eta^2\xi^2}\notag\\
                      &=&\frac{(L_{\phi_1}+L_{\phi_2})^2(H_{,\eta}^2+H_{,\xi}^2)}{K^2}>0,
\end{eqnarray}
where we have used $G_{,\eta}=-(L_{\phi_1}+L_{\phi_2})H_{,\xi}\eta\xi/2$ and $G_{,\xi}=(L_{\phi_1}+L_{\phi_2})H_{,\eta}\eta\xi/2$, which are derived from $d\chi=*_{3}dH \ (\chi_{\phi,\eta}=-H_{,\xi}\eta\xi/2,\ \chi_{\phi,\xi}=H_{,\eta}\eta\xi/2)$.

For the particles with $J=0$ and $(L_{\phi_1},L_{\phi_2})\not=(0,0)$, one of the two conditions, $G=0$, is always satisfied at least on $I_\pm$ and $I_i\ (i=1,\ldots,n-1)$ because $\chi_\phi=\pm1$ on $I_\pm$ and $\chi_\phi=2n-2i+1$ on $I_i\ (i=1,2,\ldots,n-1)$. 
Therefore, for such particles with zero energy, $U'$ has a zero local minimum at the intersection of the evanescent ergosurfaces and the axes $I_\pm$ and $I_i\ (i=1,\ldots,n-1)$. 
This is why on such surfaces, massless particles with zero energy move along stable trapped null geodesics. 
In the following sections, we will discuss it from the contours of $U'$ in the $(\eta,\xi)$ plane.

%%%%%%%%%%%%%%%%%%%%%%%%%%%%%%%%%%%%%%%%%%%%
%%%%%%%%%%%%%%%%%%%%%%%%%%%%%%%%%%%%%%%%%%%%
%%%%%%%%%%%%%%%%%%%%%%%%%%%%%%%%%%%%%%%%%%%%

%                                << >>

%%%%%%%%%%%%%%%%%%%%%%%%%%%%%%%%%%%%%%%%%%%%
%%%%%%%%%%%%%%%%%%%%%%%%%%%%%%%%%%%%%%%%%%%%
%%%%%%%%%%%%%%%%%%%%%%%%%%%%%%%%%%%%%%%%%%%%
\section{Stable bound orbits}\label{sec:SBO}
\subsection{A black lens with the topology $L(2,1)$}
In the previous work, we have seen the existence of stable bound orbits for the black lens with $L(2,1)$ topology, where we have focused on the behavior of the effective potential $U$ on the $z$ axis of ${\mathbb E}^3$ in the Gibbons-Hawking space. 
The $z$ axis is composed of the $3$ intervals: 
$I_-=\{(\eta,\xi)|\eta=0,\xi>0\}$, $I_1=\{(\eta,\xi)|\eta_1(=0)<\eta<\eta_{2},\xi=0\}$ and $I_+=\{(\eta,\xi)|\eta>\eta_2,\xi=0\}$. 
On $I_+$ and $I_-$, only the particles with the angular momenta of
 $l_{\phi_2}=0$ and $l_{\phi_1}=0$, respectively, are allowed to stay there, 
 whereas only ones with the angular momenta of the ratio $l_{\phi_1}/l_{\phi_2}=-2$ are allowed to stay on $I_1$.

\medskip
\subsubsection{$I_+$}
\begin{figure}[h]
 \begin{tabular}{cc}
 \begin{minipage}[t]{0.5\hsize}
\includegraphics[width=7cm,height=7cm]{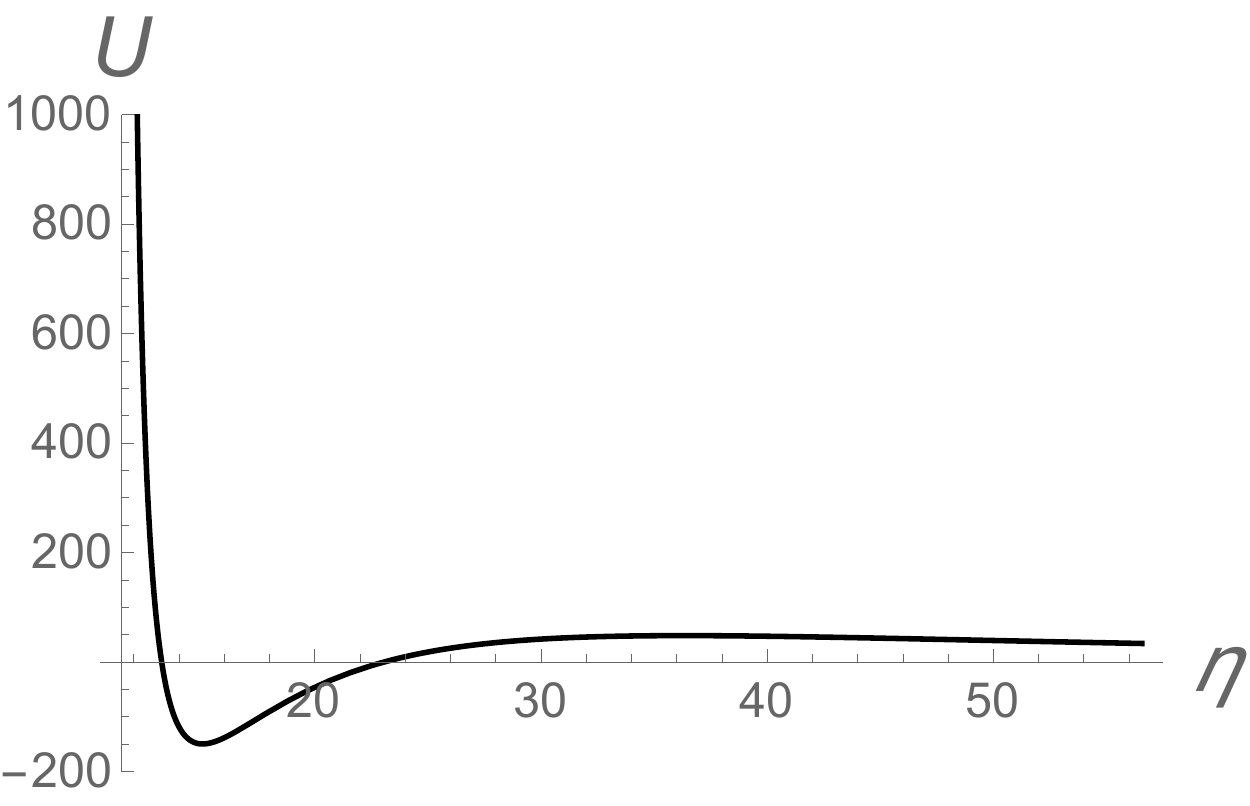}
 \end{minipage} &
 
 \begin{minipage}[t]{0.5\hsize}
\includegraphics[width=7cm,height=7cm]{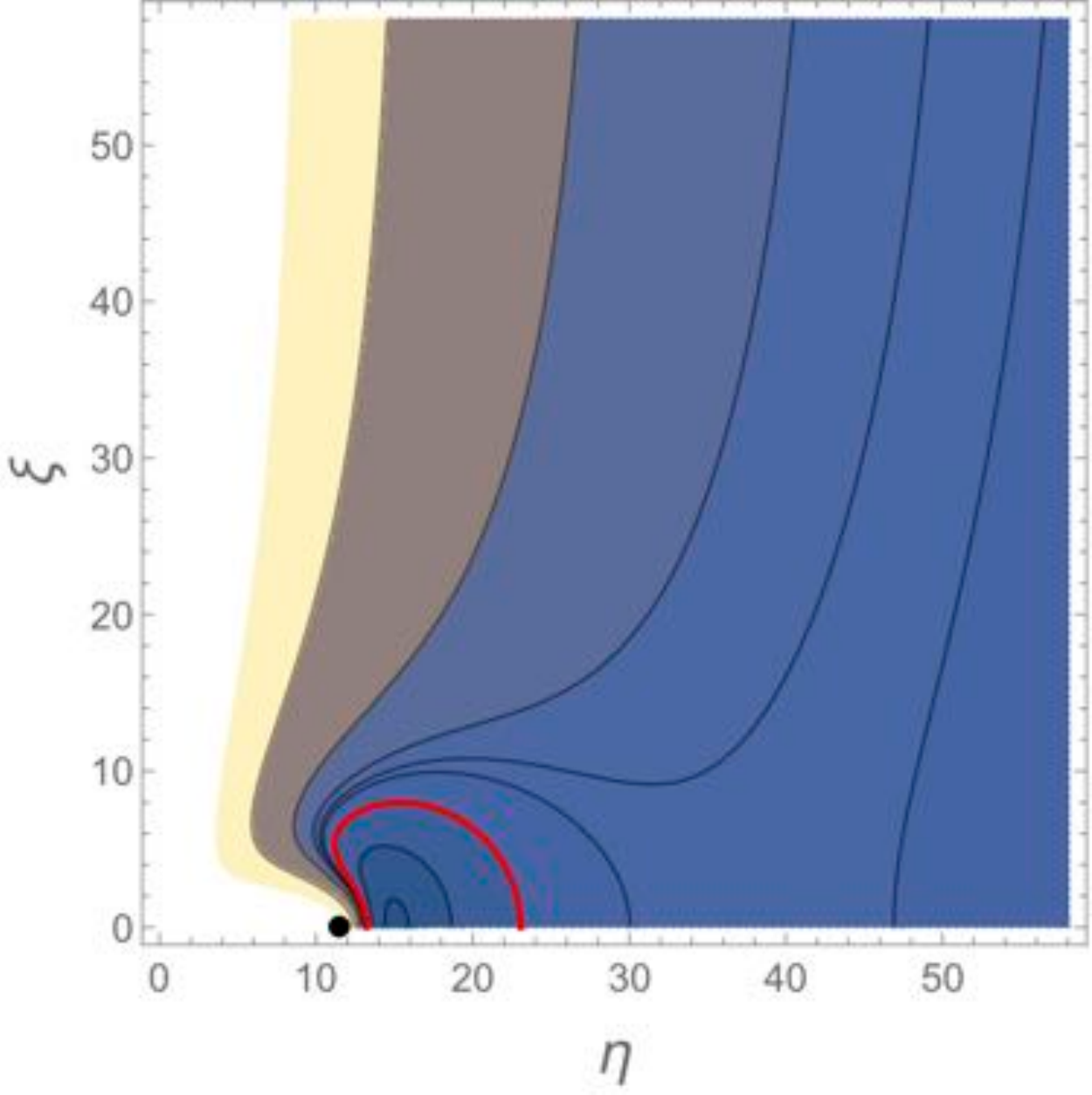}
 \end{minipage}

\end{tabular}
\caption{The left figure shows the effective potential $U$ on $I_+\ (\eta>\eta_2= 11.4\cdots,\ z>z_2= 32.8\cdots)$ for $(k_1,k_2,l_1)=(0,10,1)$ and $(l_{\phi_1},l_{\phi_2})=(-400,0)$. The right figure shows the two-dimensional plot of $U$ and the red curve corresponds to $U=0$.}
\label{fig:potential_on_I_+}\end{figure}

First, we begin to comment on the existence of stable bound orbits on $I_+$. 
Figure~\ref{fig:potential_on_I_+} shows the typical shape of the effective potential $U$ with a negative local minimum on $I_+$, where we plot $U$ under the parameter setting $(k_1,k_2,l_1)=(0,10,1)$ and the angular momenta $(l_{\phi_1},l_{\phi_2})=(-400,0)$. 
The left figure shows the shape of the effective potential on $I_+$, which is the intersection of $U$ by $\xi=0\ (\eta_2<\eta)$. 
The right figure shows the contours of $U$ for the particles with the same angular momenta. 
The red curve in this figure, which corresponds to $U=0$, separates the two-dimensional $(\eta,\xi)$-plane into two regions, the outer region $U>0$ and the inner region $U<0$. 
As can be read off from the contours, $U$ has a negative local minimum at a certain point $(\eta,\xi)=(\eta_{\mathrm{m}},0)$ on the $\eta$ axis ($z=z_{\mathrm m}$ on the $z$ axis), where $\eta_{\mathrm m}(=2\sqrt{z_{\mathrm m}})$ is a certain constant satisfying $\eta_2<\eta_{\mathrm m}$.
At the point $U_{,\eta}=U_{,\xi}=0$, ${\rm det}(H_{ij})>0$ and ${\rm Tr}(H_{ij})>0$.
Therefore, both massive and massless particles inside the red curve are stably bounded in the finite region $U\le -m^2/E^2$ and $U\le 0$, respectively. 
This means that there are stable bound orbits for massive/massless particles.

\medskip
Furthermore, we consider whether there can exist such stable bound orbits at infinity $z\to\infty\ (\eta \to\infty,\ \xi=0)$ on $I_+$. 
To this end, let us expand the potential $U$ at $z\to \infty$ on $I_+$ as
\begin{eqnarray}
U\simeq -1+\frac{U^{(1)}_\infty}{z}+\frac{U^{(2)}_\infty}{z^2},
\end{eqnarray}
where the constants $U^{(1)}_\infty$ and $U^{(2)}_\infty$ are denoted by
\begin{eqnarray}
U^{(1)}_\infty=\frac{\l_{\phi_1}^2-16k_2^2-8l_1}{4},\quad 
U^{(2)}_\infty=\frac{-(14k_2^2+3l_1)\l_{\phi_1}^2+96k_2^3\l_{\phi_1}-160k_2^4-24l_1^2}{24}.
\end{eqnarray}
After simple computations, we find that $0<-U^{(1)}_\infty\ll 1$ and $U^{(2)}_\infty>0$ are the necessary conditions that $U$ has a local minimum at infinity because the local minimum is at $z\simeq -2U^{(2)}_\infty/U^{(1)}_\infty$. 
Noting that the condition $0<-U^{(1)}_\infty\ll 1$ can be denoted by $\l_{\phi_1}=\pm 2\sqrt{2(2k_2^2+l_1)}\mp \epsilon\ (0<\epsilon\ll 1)$, we can confirm that the leading term of $U^{(2)}_\infty$ can be written as
\begin{eqnarray}
U^{(2)}_\infty\simeq
-\frac{2}{3}(24k_2^4+10l_1k_2^2+3l_1^2\mp12k_2^3\sqrt{2(2k_2^2+l_1)})+{\cal O}(\epsilon).
\end{eqnarray}
From the leading terms, we find that $U^{(2)}_\infty <0$, which cannot satisfy the other condition $U^{(2)}_\infty>0$. 
As a result, there exist no stable bound orbits in the asymptotic region $z\to \infty$, at least, on the $z$ axis. 
This suggests that there exists the outermost stable circular orbit as the boundary of stable circular orbits.

\subsubsection{$I_-$}
\begin{figure}[h]
 \begin{tabular}{cc}
 \begin{minipage}[t]{0.5\hsize}
\includegraphics[width=7cm,height=7cm]{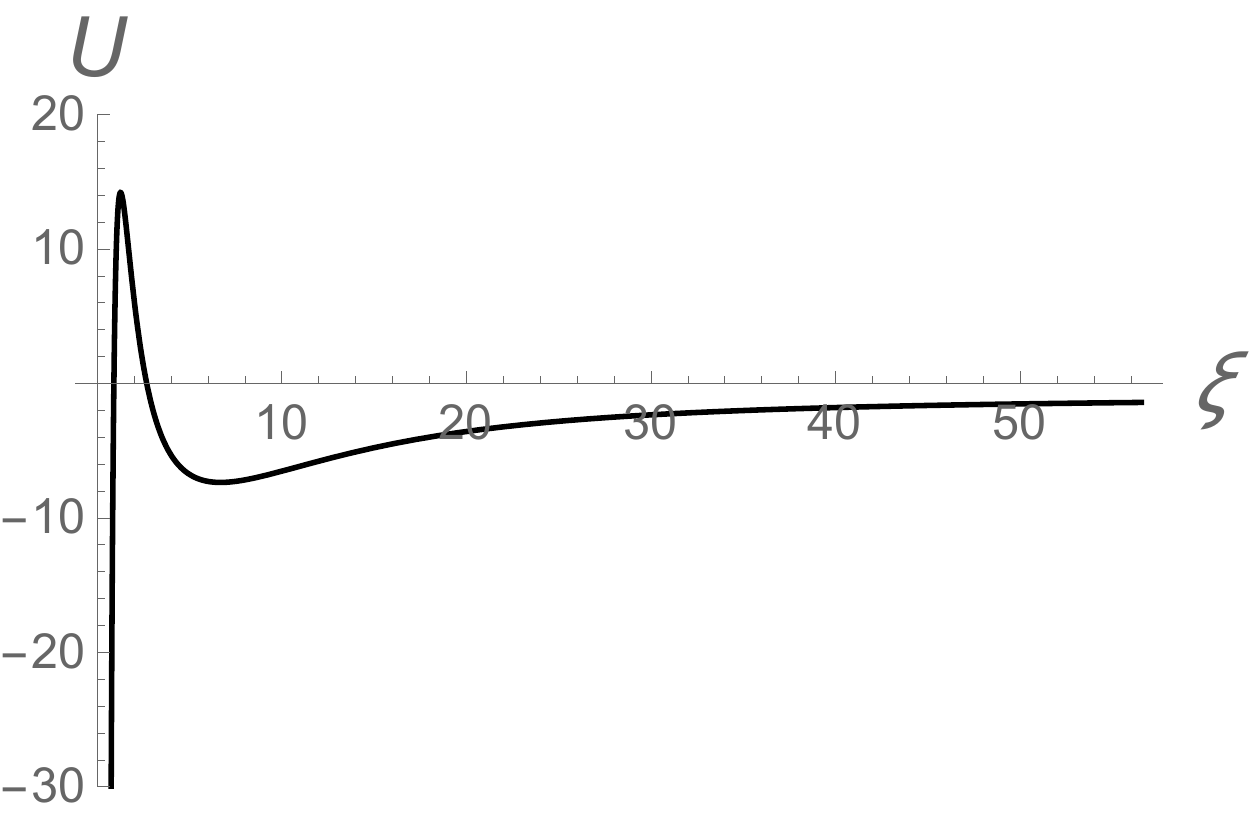}
 \end{minipage}&
 
 \begin{minipage}[t]{0.5\hsize}
\includegraphics[width=7cm,height=7cm]{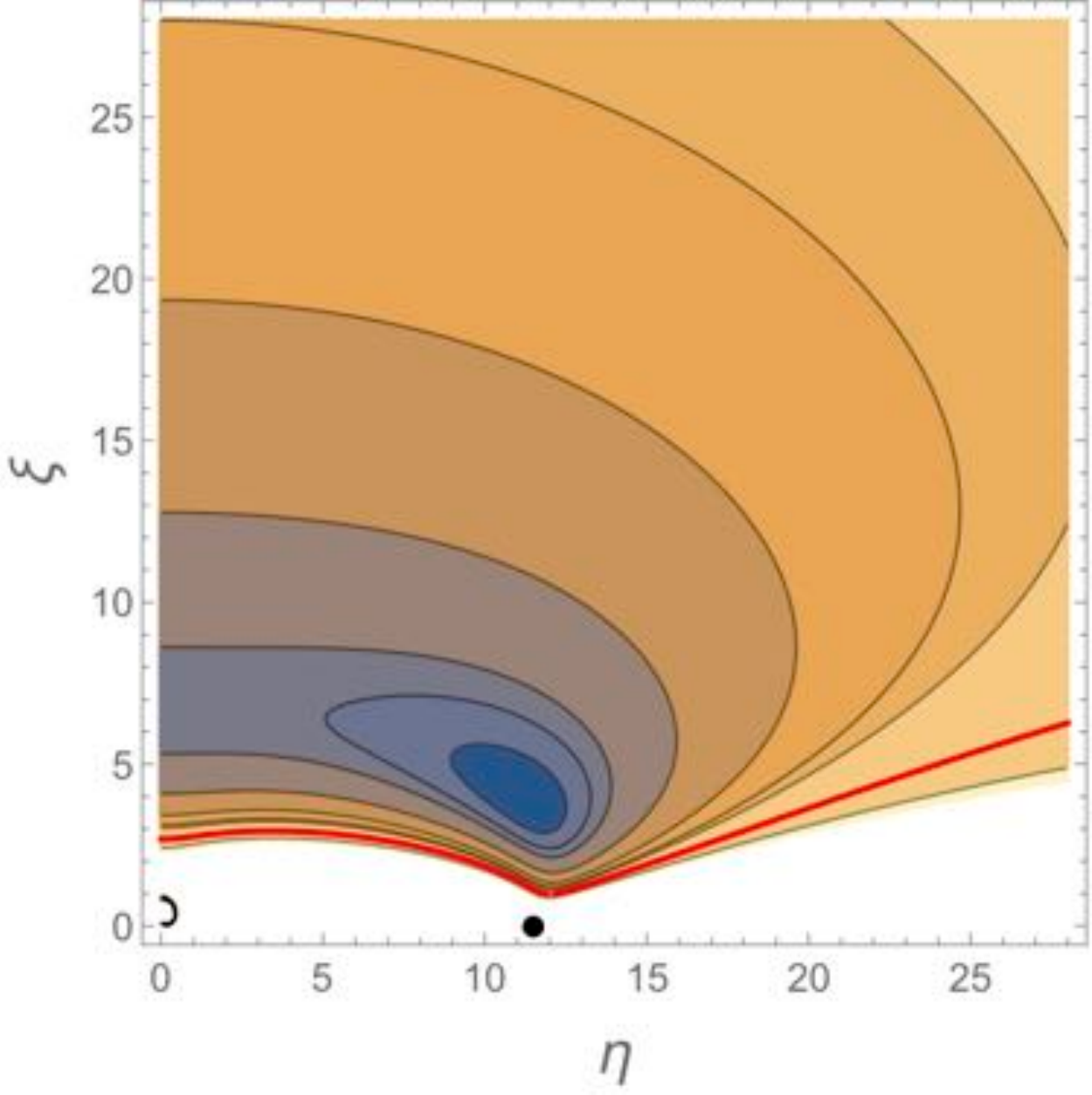}
 \end{minipage}

\end{tabular}

\caption{The left figure shows the effective potential $U$ on $I_-\ (\eta=0,\ \xi>0)$ for $(k_1,k_2,l_1)=(0,10,1)$ and $(l_{\phi_1},l_{\phi_2})=(0,16)$. The right figure shows the corresponding two-dimensional plot of $U$.}

\label{fig:potential_on_I_-}
\end{figure}

Next, we focus on the potential on $I_-$, where the particles with angular momenta $l_{\phi_1}=0,\ l_{\phi_2}\not=0$ can stay. 
The left graph in Fig.~\ref{fig:potential_on_I_-} shows the intersection of $U$ by $\eta=0$ for $(l_{\phi_1},l_{\phi_2})= (0,16)$ in the same choice of the parameters $(k_1,k_2,l_1)$. 
The right figure shows the contours of $U$. 
At the horizon $(\eta,\xi)=(0,0)$, $U$ diverges to $-\infty$ due to the gravitational force. 
The potential $U$ seems to have a local minimum but this is not the case because the Hessian gets negative. 
This shows that there exist unbounded orbits of massless particles on $I_-$.
As seen from the right figure, a negative local minimum exist at the place far from rather the axis than on it. 
For massless particles, there are no stable bound orbits because the red curve $U=0$ is not closed, whereas for massive particles with the mass $m$, there are stable bound orbits within the region $U\le -m^2/E^2$.

\medskip
Let us discuss whether there can exist stable bound orbits at the other infinity $z\to-\infty\ (\eta=0,\xi\to\infty)$ on the $z$ axis. 
To do so, we expand the potential $U$ at $z\to -\infty$ as
\begin{eqnarray}
U\simeq -1+\frac{U^{(1)}_{-\infty}}{z}+\frac{U^{(2)}_{-\infty}}{z^2},
\end{eqnarray}
where the constants $U^{(1)}_{-\infty}$ and $U^{(2)}_{-\infty}$ are denoted by
\begin{eqnarray}
U^{(1)}_{-\infty}&=&\frac{-\l_{\phi_2}^2+16k_2^2+8l_1}{4},\\
U^{(2)}_{-\infty}&=&\frac{-(10k_2^2+9l_1)\l_{\phi_2}^2+(48k_2^3+72k_2l_1)\l_{\phi_2}-32k_2^4-192k_2^2l_1-24l_1^2}{24}.
\end{eqnarray}
The inequalities $0<U^{(1)}_{-\infty}\ll 1$ and $U^{(2)}_{-\infty}>0$ are necessary for $U$ to have a local minimum at the infinity $z\to -\infty$. 
The condition $0<U^{(1)}_{-\infty}\ll 1$ can be denoted by $\l_{\phi_2}=\pm 2\sqrt{2(2k_2^2+l_1)}\mp \epsilon\ (0<\epsilon\ll 1)$, and the leading term of $U^{(2)}_{-\infty}$ can be written as
\begin{eqnarray}
U^{(2)}_{-\infty}\simeq-\frac{2}{3}\left(12k_2^4+26l_1k_2^2+6l_1^2\mp3(2k_2^3+3k_2l_1)\sqrt{2(2k_2^2+l_1)}\right)+{\cal O}(\epsilon).
\end{eqnarray}
From the leading terms, we find that $U^{(2)}_{-\infty} <0$ for any parameter $l_1,\ k_2$, which cannot satisfy the necessary condition for the existence of stable bound orbits in the asymptotic region $z\to -\infty$, at least, on the $z$ axis. 
 
 \medskip
Moreover, let us expand $U$ near the horizon $z= 0$ as
 \begin{eqnarray}
U\simeq -\frac{l_1^2}{z^2}+\frac{2l_1(8k_2^2-3l_1)}{(2k_2^2-3l_1)z}+{\cal O}(1),
\end{eqnarray}
and then we find that the first and second terms are negative, which is due to the effect of the attraction by the horizon. 
From Fig.~\ref{fig:potential_on_I_-}, 
$U$ does not make a local minimum on the $z$ axis.

\subsubsection{$I_1$}
\begin{figure}[h]
 \begin{tabular}{cc}
 
 \begin{minipage}[t]{0.5\hsize}
\includegraphics[width=7cm,height=7cm]{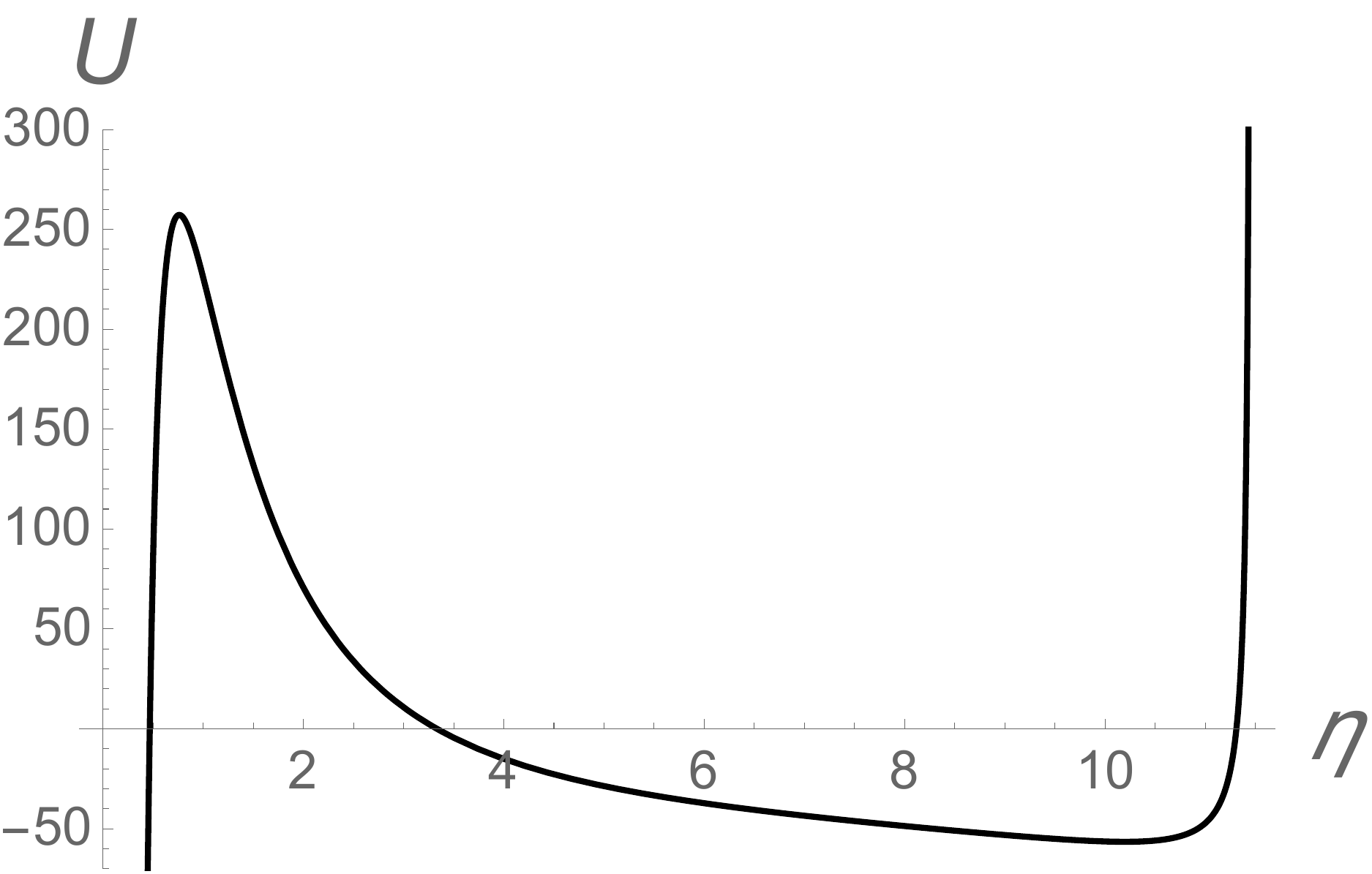}
 \end{minipage} & 
 \begin{minipage}[t]{0.5\hsize}
\includegraphics[width=7cm,height=7cm]{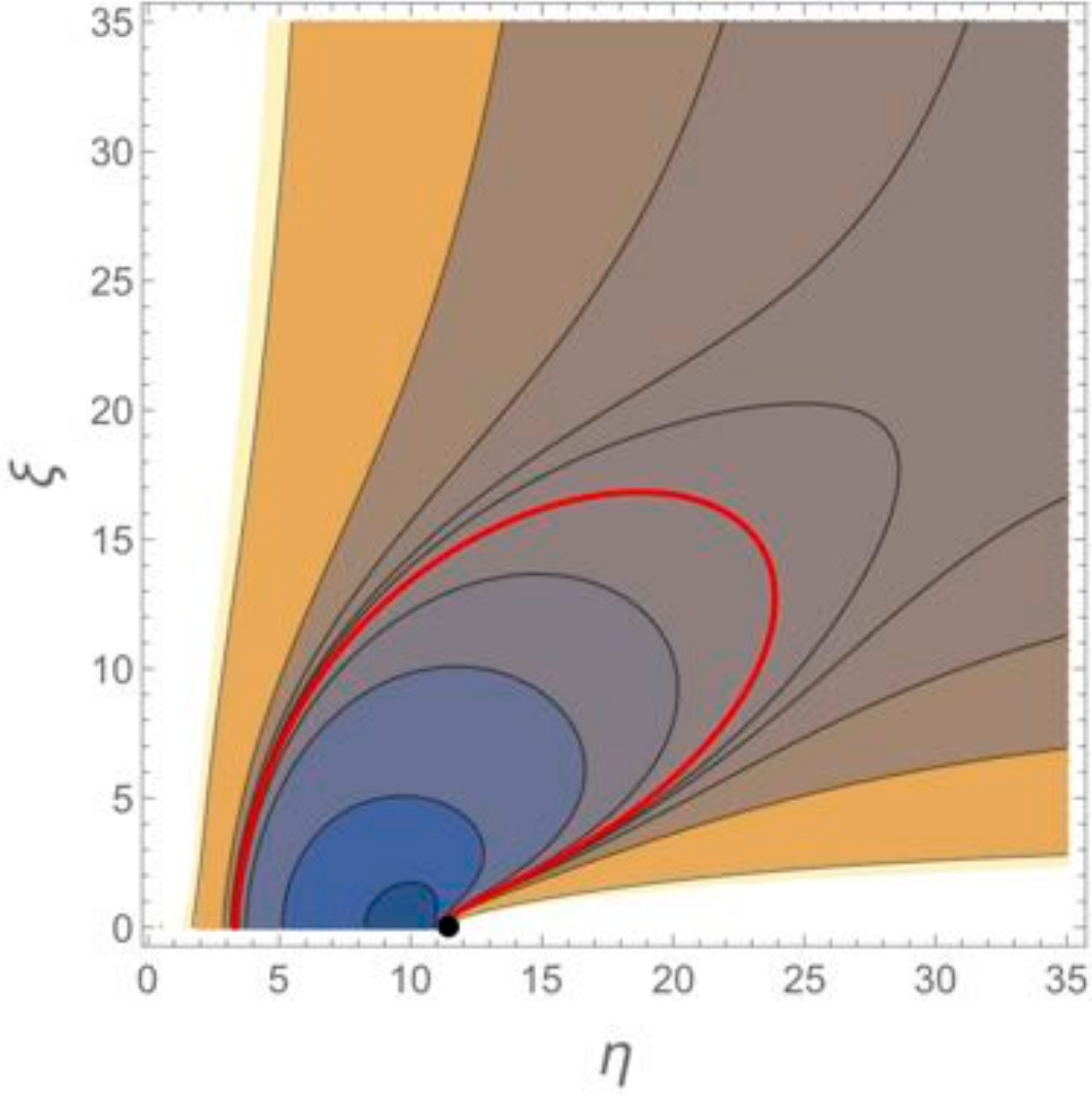}
 \end{minipage} 
 
 \end{tabular}
\caption{
The left figure shows the effective potential $U$ on $I_1\ (0<\eta<\eta_2=11.4\cdots,\ \xi=0)$ for $(k_1,k_2,l_1)=(0,10,1)$ and $(l_{\phi_1},l_{\phi_2})=(-72,36)$. The right figure shows the corresponding two-dimensional plot of $U$.}
\label{fig:potential_on_I_1}
\end{figure}

Finally, we consider the potential $U$ on the interval $I_1$, where the particles with the angular momenta of the special ratio of $l_{\phi_1}/l_{\phi_2}=-2$ are allowed to stay. 
Figure~\ref{fig:potential_on_I_1} shows the typical behavior of the potential $U$ with a local minimum for the same set of parameters $(k_1,k_2,l_1)=(0,10,1)$ and $(l_{\phi_1},l_{\phi_2})=(36,-18)$. 
Near the horizon at $(\eta,\xi)=(0,0)$, $U$ increases by the effect of the centrifugal force, while as closer to the horizon, the potential diverges to $-\infty$ by the stronger effect of the gravitational force.
On the other hand, near the other center $(\eta,\xi)=(\eta_2,0)$, due to the effect of the centrifugal force, 
the potential again increases rapidly, and then diverges to $\infty$.
As a result, there is necessarily a negative local minimum somewhere between the horizon $(\eta,\xi)=(0,0)$ and the center $(\eta,\xi)=(\eta_2,0)$ on the $\eta$ axis. 
This is the reason why there exist stable bound orbits of massless particles as well as massive particles in a black lens spacetime.

\subsection{A black lens with the topology $L(3,1)$}
Next, let us consider the black lens with the topology of $L(3,1)$. 
The $z$ axis of the Gibbons-Hawking space is composed of the $4$ intervals: 
$I_-=\{(\eta,\xi)|\eta=0,\ \xi>0\}$, $I_1=\{(\eta,\xi)|0<\eta<\eta_{2},\ \xi=0\}$, $I_2=\{(\eta,\xi)|\eta_2<\eta<\eta_{3},\ \xi=0\}$ and $I_+=\{(\eta,\xi)|\eta>\eta_3,\ \xi=0\}$. 
On $I_+$ and $I_-$, only the particles with the angular momenta of
 $l_{\phi_2}=0$ and $l_{\phi_1}=0$, respectively, are allowed to stay there, 
 whereas only ones with the angular momenta of the ratio $l_{\phi_1}/l_{\phi_2}=-(4-i)/(3-i)$ are allowed to stay on $I_i\ (i=1,2)$.

\subsubsection{$I_+$}
\begin{figure}[h]
 \begin{tabular}{cc}
 \begin{minipage}[t]{0.5\hsize}
\includegraphics[width=7cm,height=7cm]{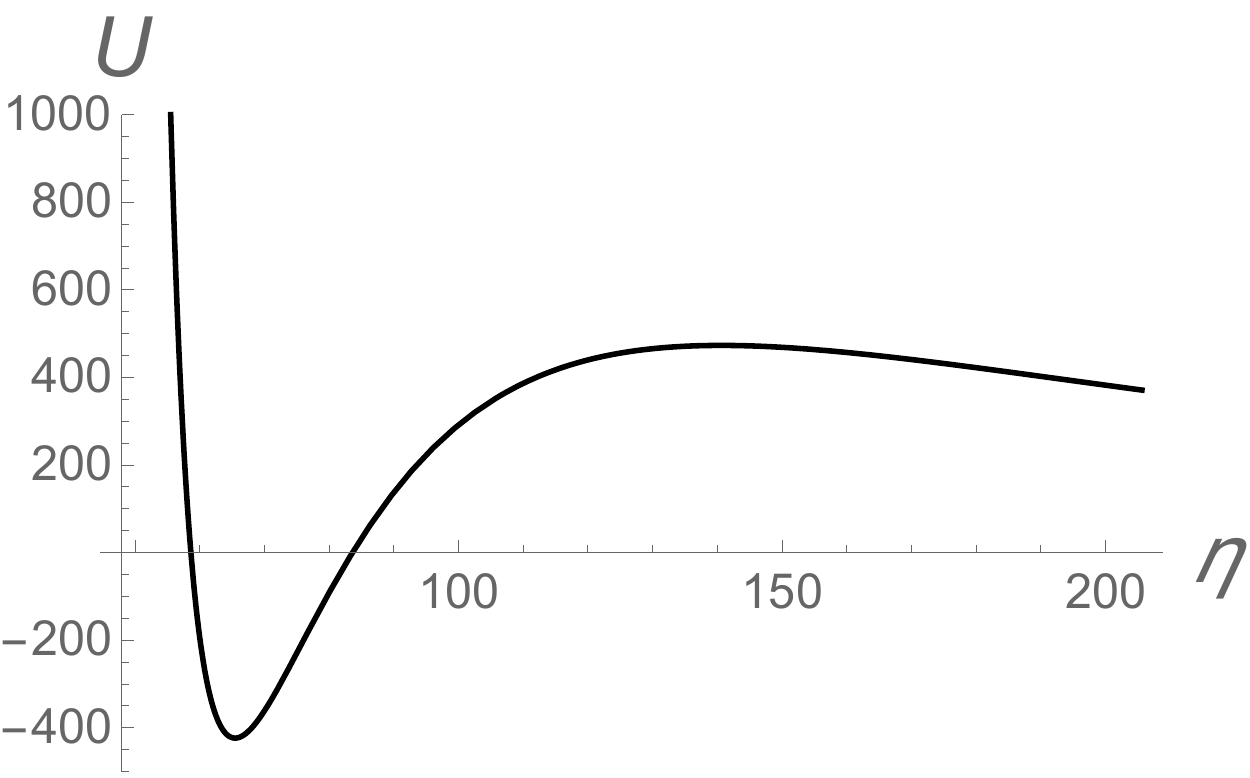}
 \end{minipage} & 
 
 \begin{minipage}[t]{0.5\hsize}
\includegraphics[width=7cm,height=7cm]{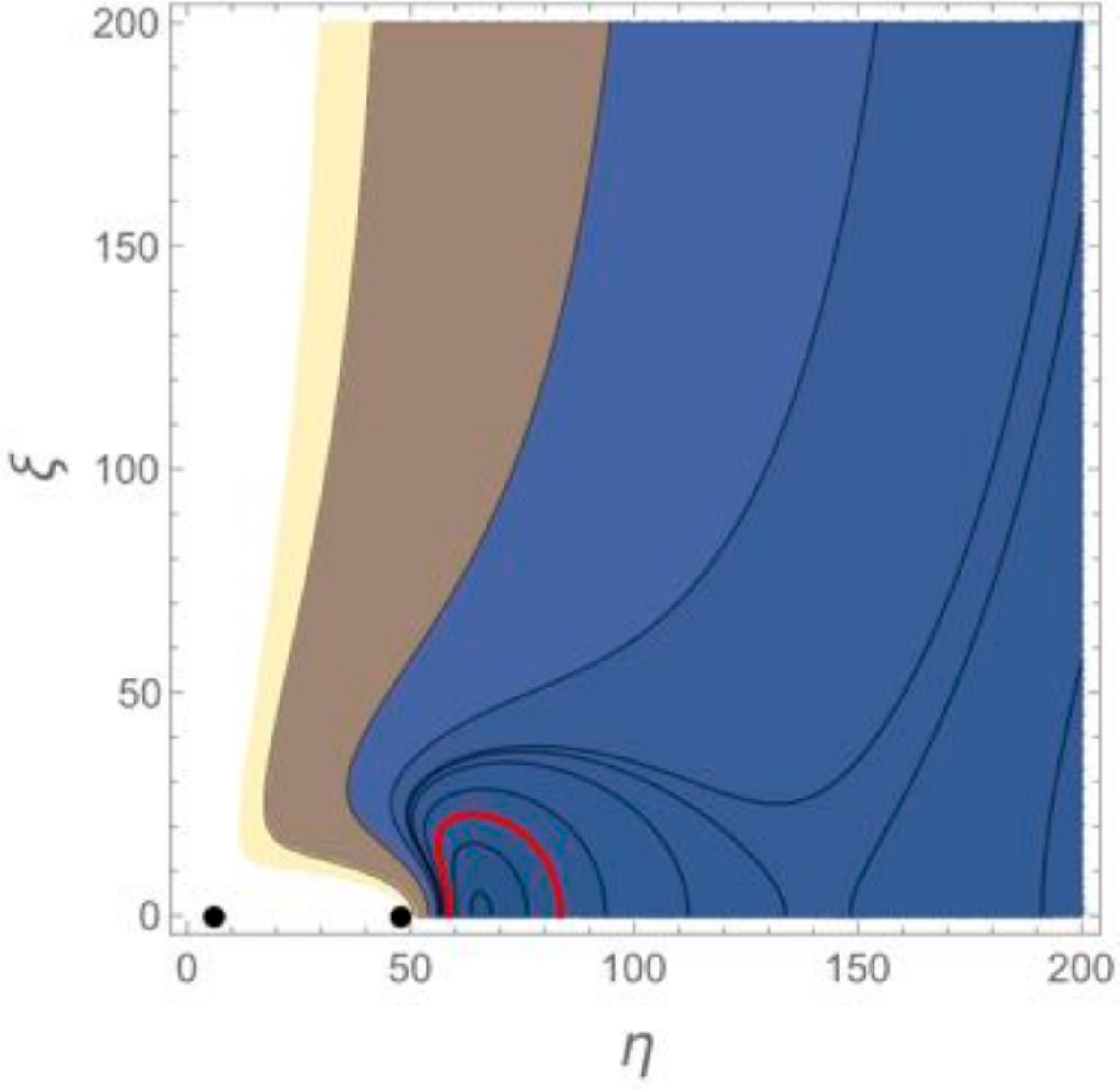}
 \end{minipage}  
 \end{tabular}
\caption{The left figure shows the potential $U$ for the particles with the angular momenta of $(l_{\phi_1},l_{\phi_2})=(5000,0)$ for $(k_1,k_2,k_3,l_1)=(0,10,-50,1)$. 
The right figure shows the contours of $U$ for the particles with the same angular momenta. 
The red closed curve $U=0$ separates the two-dimensional $(\eta,\xi)$-plane into two regions $U>0$ and $U<0$. 
The two black points in the right figure denote the points $(\eta_2,0)$ and $(\eta_3,0)$, where $\eta_2=6.36\cdots$, $\eta_3=47.91\cdots$ and the origin $(0,0)$ corresponds to the horizon.
}
\label{fig:L31_I+}

\label{fig:I+z3}
\end{figure}

\medskip
First, let us see the shape of the potential $U$ on $I_+$. 
The left figure of Fig.~\ref{fig:L31_I+} shows the shape of the typical effective potential with a negative local minimum on the $z$ axis for the particles with the angular momenta of $(l_{\phi_1},l_{\phi_2})=(5000,0)$ for $(k_1,k_2,k_3,l_1)=(0,10,-50,1)$. 
The right figure shows the contours of $U$ for the particles with the same angular momenta. 
The closed red curve in this figure, which corresponds to $U=0$, separates the two-dimensional $(\eta,\xi)$-plane into two regions, the outer region $U>0$ and the inner region $U<0$. 
As can be read off from the contours, $U$ has a negative local minimum at a certain point $(\eta,\xi)=(\eta_{\mathrm m},0)$ on the $\eta$ axis ($z_{\mathrm m}$ on the $z$ axis), where $\eta_{\mathrm m} (=2\sqrt{z_{\mathrm m}})$ is a certain constant satisfying $\eta_2<\eta_{\mathrm m}$.
At the point $U_{,\eta}=U_{,\xi}=0$, ${\rm det}(H_{ij})>0$ and ${\rm Tr}(H_{ij})>0$.
Therefore, both massive and massless particles inside the red curve are stably bounded in the finite region $U\le -m^2/E^2$ and $U\le 0$, respectively. 
This means that there are stable bound orbits for massive/massless particles. 

\medskip
Moreover, let us consider whether there exist stable bound orbits at infinity $z\to\infty$. 
To this end, let us expand the potential $U$ at $z\to \infty$ as
\begin{eqnarray}
U\simeq -1+\frac{U^{(1)}_\infty}{z}+\frac{U^{(2)}_\infty}{z^2},
\end{eqnarray}
where 
\begin{eqnarray}
U^{(1)}_\infty=\frac{l_{\phi_1}^2-16(k_2^2+k_2k_3+k_3^2)-8l_1}{4},
\end{eqnarray}
and $U^{(2)}_\infty$ can be written as a certain quadratic equation for $l_{\phi_1}$. 
It turns out from this asymptotic behavior that the inequalities $0<-U^{(1)}_\infty\ll 1$ and $U^{(2)}_\infty>0$ are required so that $U$ has a local minimum at infinity because the local minimum is at $z\simeq-2U^{(2)}_\infty/U^{(1)}_\infty$. 
The former condition $0<-U_\infty^{(1)}\ll 1$ can be denoted in the form of $l_{\phi_1}=\pm 2\sqrt{4(k_2^2+k_2k_3+k_3^2)+2l_1}\mp \epsilon\ (0<\epsilon\ll 1)$, under whose condition the sign of $U^{(2)}_\infty$ is determined by the leading term of $U^{(2)}_\infty$ in the expansion $U^{(2)}_\infty\simeq U^{(2)}_{\infty 0}+{\cal O}(\epsilon)$. 
However, we can see numerically
that the region $U^{(2)}_{\infty0} >0$ does not overlap the parameter region ${\cal D}$, which means that there are no stable bound orbits on the $z$ axis in the asymptotic region $z\to \infty$.

Next we study the existence of stable bound orbits near the endpoint $z=z_3$ of $I_+$, near which the potential behaves as
\begin{eqnarray}
U\simeq \frac{U^{(-1)}_{3+}}{z-z_3}+U^{(0)}_{3+}+U^{(1)}_{3+}(z-z_3),
\end{eqnarray}
where 
\begin{eqnarray}
U^{(-1)}_{3+}=-\frac{[(k_2-k_3)^3z_3+((l_{\phi_1}-3k_2-6k_3)z_3-3l_1k_3+3k_3^3)z_{32}]^2}{4z_3z_{32}[(k_2-k_3)^2z_3+(l_1-3k_3^2+z_3)z_{32}]}.
\end{eqnarray}
We find numerically that $U^{(-1)}_{3+}>0$ is always satisfied within the parameter region ${\cal D}$, so that the potential diverges to $\infty$ at $z=z_3$.
Moreover, when $0<U^{(-1)}_{3+}\ll 1$, $U^{(1)}_{3+}>0$ and $U^{(0)}_{3+}+2\sqrt{U^{(-1)}_{3+}U^{(1)}_{3+}}<0$, 
$U$ has a negative local minimum near $z=z_3$ on the $z$ axis.\\

\begin{figure}[h]
 \begin{tabular}{cc}
 \begin{minipage}[t]{0.5\hsize}
\includegraphics[width=7cm,height=7cm]{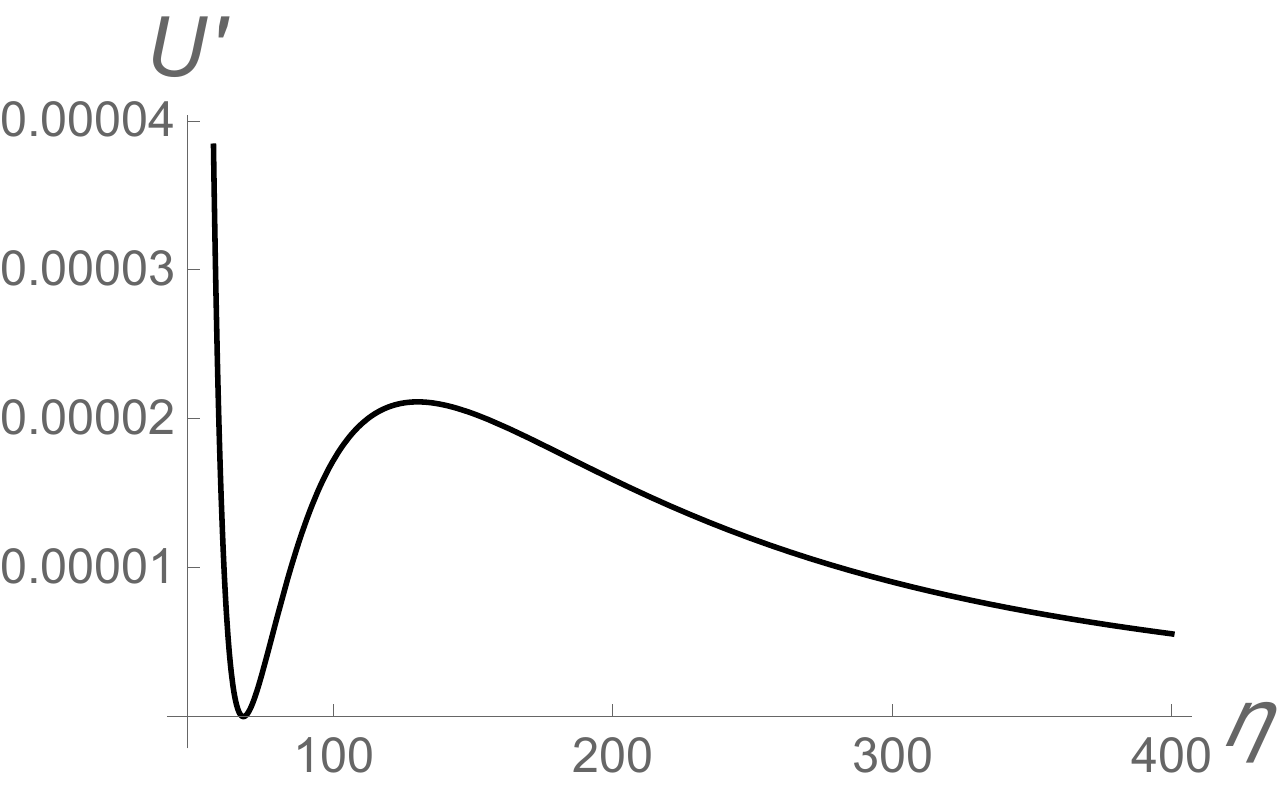}
 \end{minipage} & 
 
 \begin{minipage}[t]{0.5\hsize}
\includegraphics[width=7cm,height=7cm]{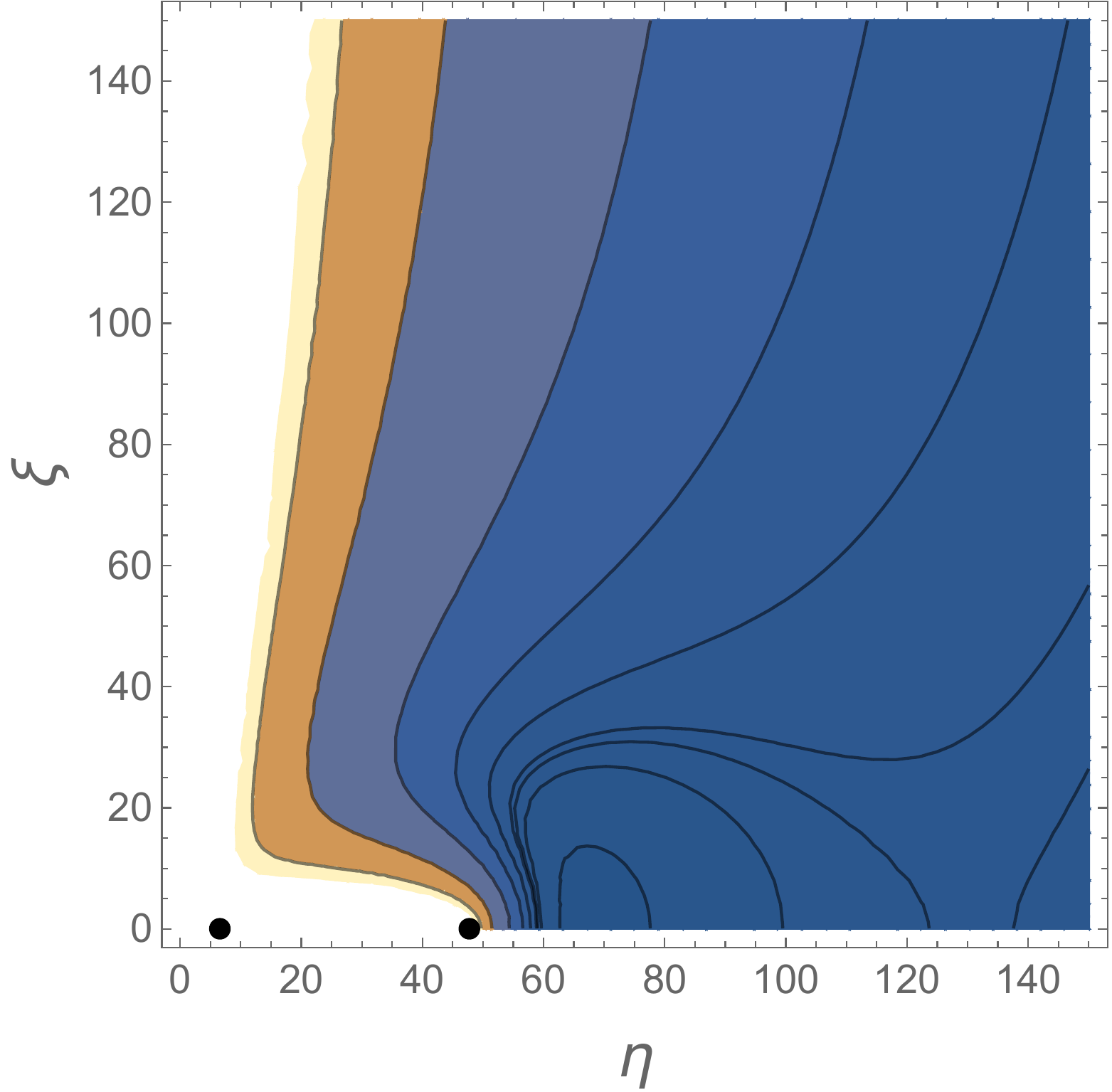}
 \end{minipage} 
 \end{tabular}
\caption{The left figure shows the potential $U'$ under the parameter setting $(k_1,k_2,k_3,l_1)=(0,10,-50,1)$ and angular momenta $(L_{\phi_1},L_{\phi_2})=(1,0)$. 
The right figure shows the contours of $U'$ for the particles with the same angular momenta. The potential has a local minimum, whose values are zero on the evanescent ergosurfaces. }
\label{fig:U'I+}
\end{figure}

Let us consider the geodesic motion of the massless particles with zero-energy limit $E\to 0$, in which case it is more convenient to use the potential $U'$ rather than $U$. 
In Fig.~\ref{fig:U'I+}, the left figure shows the typical shape of the effective potential $U'$ for the particles with the angular momenta $L_{\phi_2}\not=0$, where we put $(k_1,k_2,k_3,l_1)=(0,10,-50,1)$ and $(L_{\phi_1},L_{\phi_2})=(1,0)$. 
The right figure shows the contours of $U'$ for the particles with the same angular momenta. 
The potential $U'$ has a zero local minimum on the evanescent ergosurface, where massless particles with zero energy are stably trapped.

\subsubsection{$I_-$}
\begin{figure}[h]
 \begin{tabular}{cc}
 \begin{minipage}[t]{0.5\hsize}
\includegraphics[width=7cm,height=7cm]{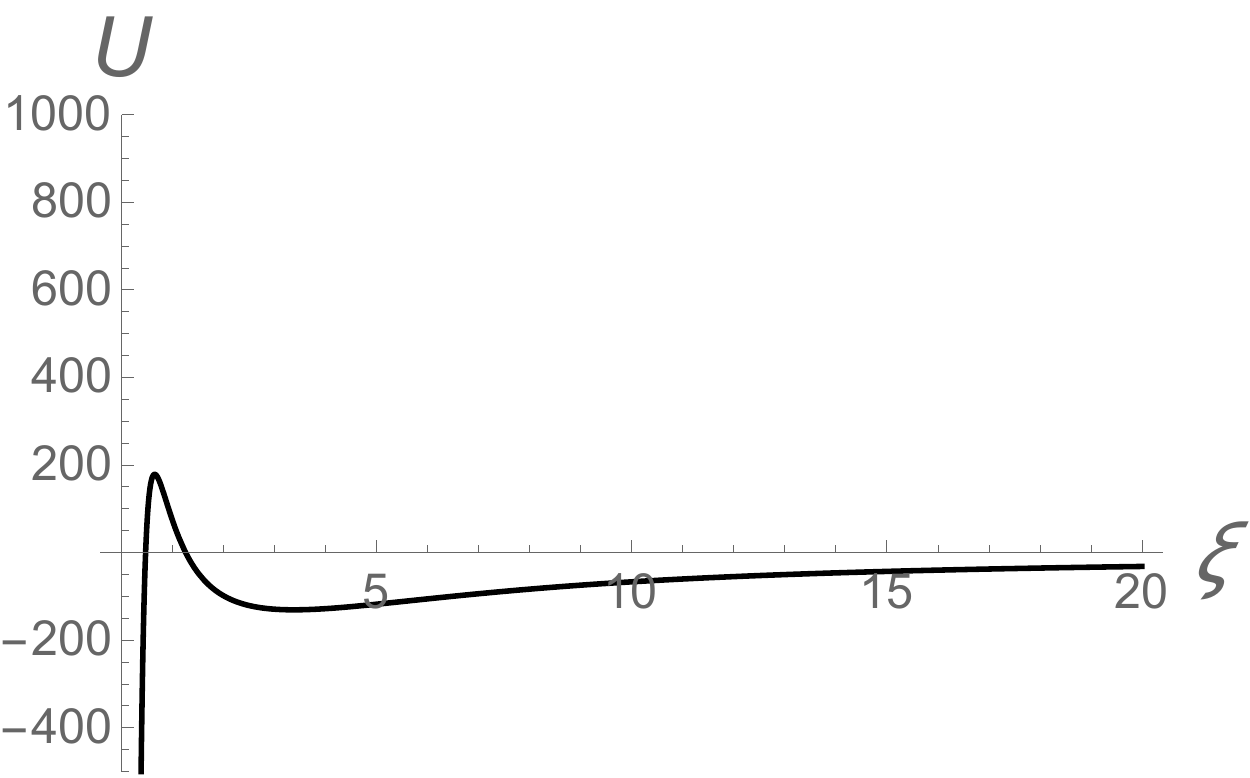}
 \end{minipage} & 
 
 \begin{minipage}[t]{0.5\hsize}
\includegraphics[width=7cm,height=7cm]{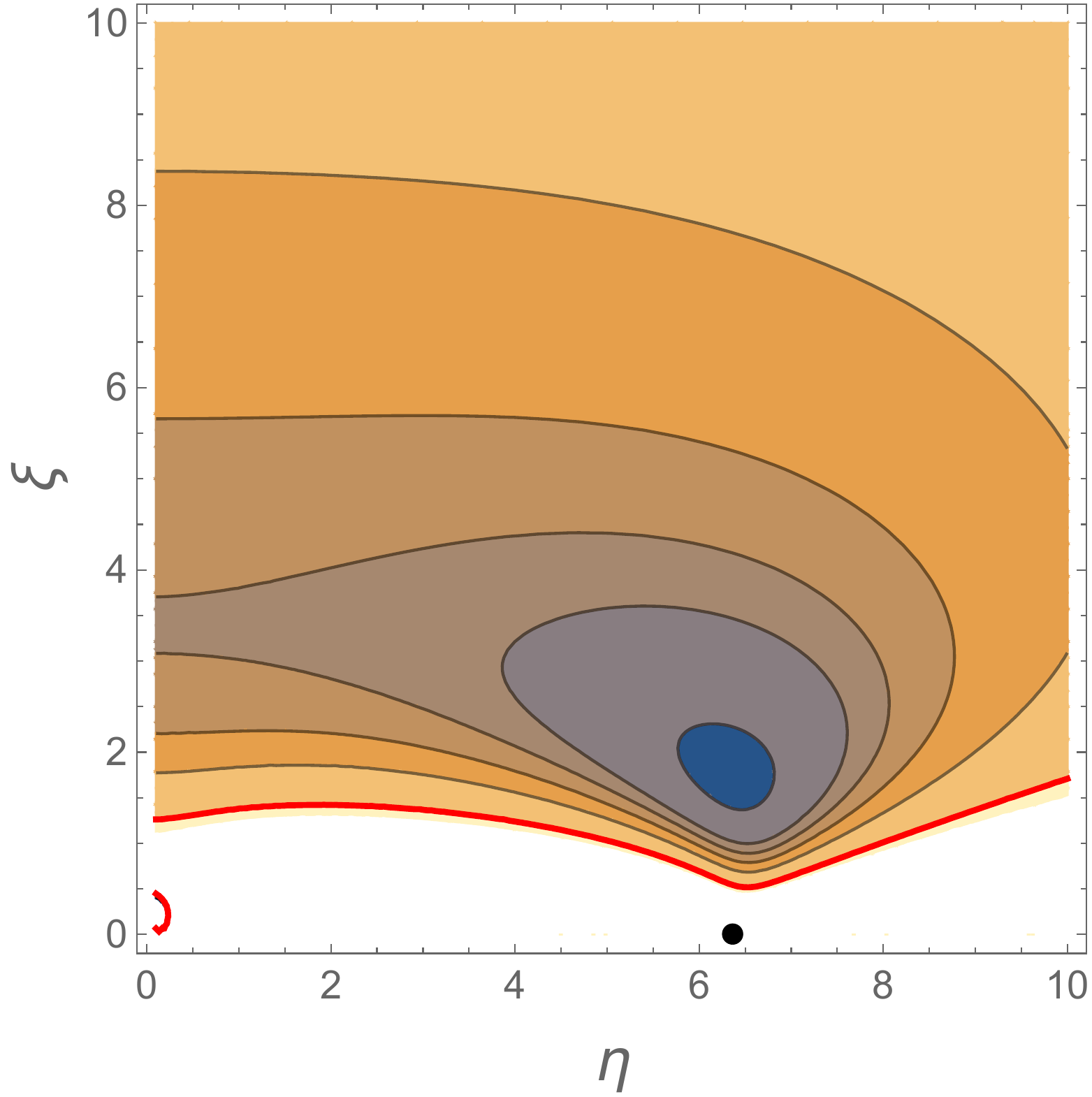}
 \end{minipage} 
 \end{tabular}
\caption{The left figure shows the effective potential $U$ for the particles with the angular momenta of $(l_{\phi_1},l_{\phi_2})=(0,50)$ for $(k_1,k_2,k_3,l_1)=(0,10,-50,1)$. 
The right figure shows the contours of $U$ for the particles with the same angular momenta. 
The red curve corresponding to $U=0$ is not closed. }

\label{fig:L31_I-}
\end{figure}

Next, let us see the shapes of the effective potential on $I_-$. 
The left figure of Fig.~\ref{fig:L31_I-} shows the typical shape of the potential $U$ with a local minimum, where this corresponds to the particles with the angular momenta of $(l_{\phi_1},l_{\phi_2})=(0,50)$ for $(k_1,k_2,k_3,l_1)=(0,10,-50,1)$. 
The right figure shows the contours of $U$ for the particles with the same angular momenta. 
The red curve corresponding to $U=0$ is not closed. 
The potential $U$ does not have a local minimum on the $\xi$ axis (on the $z$ axis corresponding to $z<0$), but as is seen from the right contour plots of $U$, there are the closed contours $U=U_0\ (U_0<0)$ which cross the $\xi$ axis. 
Therefore, massive particles crossing the $\xi$ axis inside the region $U\leq U_0$ turn out to be stably bounded.
Furthermore, massive particles can also stably bounded inside closed contours apart from the $\xi$ axis.

We discuss the existence of stable bound orbits at the infinity $z\to-\infty$ on the $z$ axis. 
Similarly we expand the potential $U$ at $z\to -\infty$ as
\begin{eqnarray}
U\simeq -1+\frac{U^{(1)}_{-\infty}}{z}+\frac{U^{(2)}_{-\infty}}{z^2},
\end{eqnarray}
where 
\begin{eqnarray}
U^{(1)}_{-\infty}=\frac{-l_{\phi_2}^2+16(k_2^2+k_2k_3+k_3^2)+8l_1}{4},
\end{eqnarray}
and $U^{(2)}_{-\infty}$ is a certain quadratic equation for $l_{\phi_2}$.
The inequalities $0<U^{(1)}_{-\infty}\ll 1$ and $U^{(2)}_{-\infty}>0$ are the necessary conditions for $U$ to have a local minimum at the infinity. 
The condition $0<U^{(1)}_{-\infty}\ll 1$ can be denoted by $l_{\phi_2}=\pm 2\sqrt{2(2k_2^2+2k_2l_3+2k_3^2+l_1)}\mp \epsilon\ (0<\epsilon\ll 1)$, then the leading term of $U^{(2)}_{-\infty}$ can be written as $U^{(2)}_{-\infty}\simeq U^{(2)}_{-\infty 0}+{\cal O}(\epsilon)$.
However, we can see numarically that the region $U^{(2)}_{-\infty 0} >0$ does not has an overlap region with the parameter region ${\cal D}$, which means that there are no stable bound orbits in the asymptotic region $z\to -\infty$ on the $z$ axis.

Near the horizon $z=0$, $U$ behaves as
\begin{eqnarray}
U\simeq -\frac{l_1^2}{z^2}+\frac{2l_1(z_2z_3+k_2^2z_3+k_3^2z_2)}{z_2z_3z}.
\end{eqnarray}
The negativity of the first and second terms makes $U$ diverge to $-\infty$ at the horizon $z=z_1(=0)$, which is a pure effect of gravity.

\begin{figure}[h]
 \begin{tabular}{cc}
\begin{minipage}[t]{0.5\hsize}
\includegraphics[width=7cm,height=7cm]{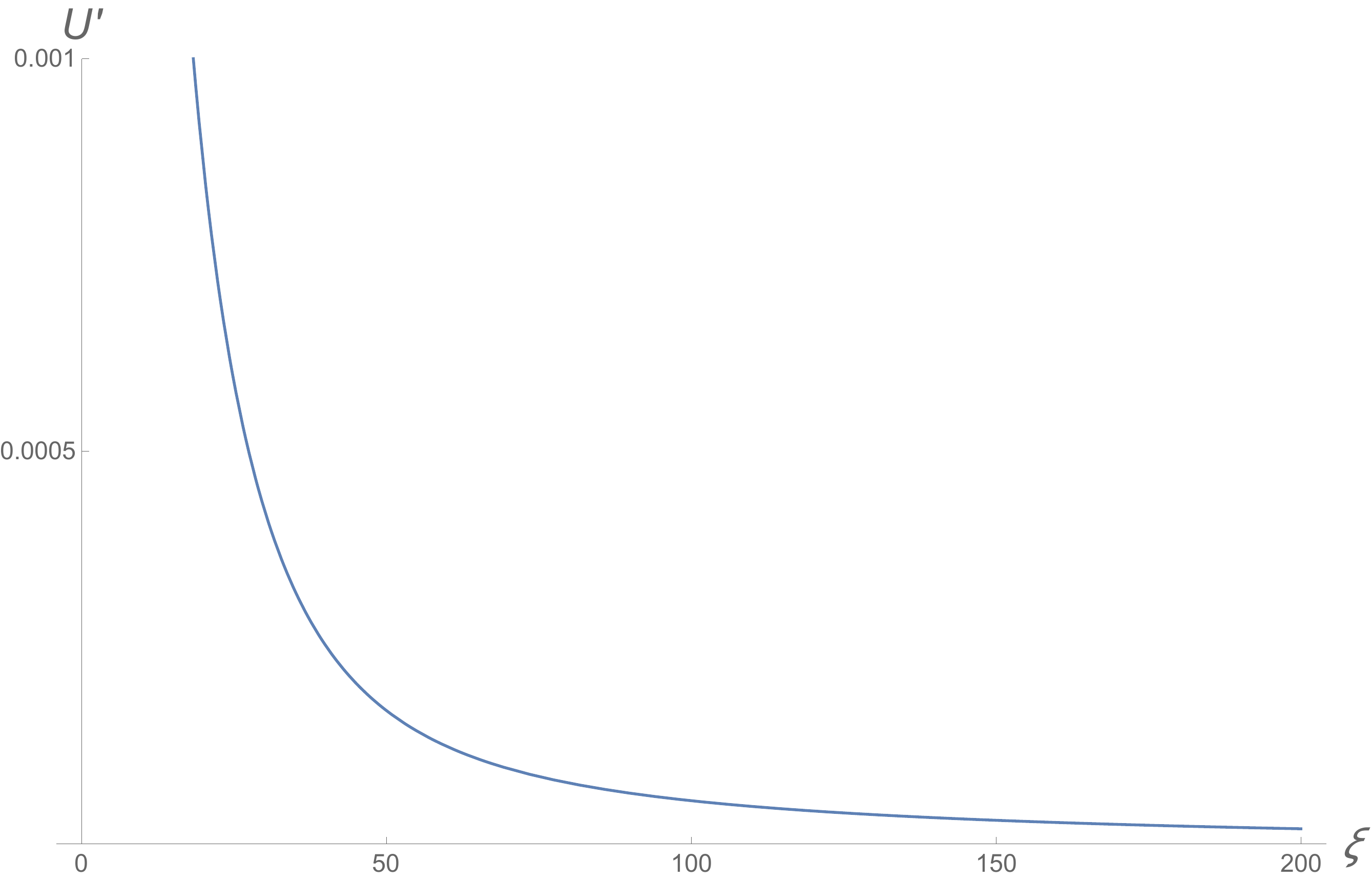}
 \end{minipage} & 

 \begin{minipage}[t]{0.5\hsize}
\includegraphics[width=7cm,height=7cm]{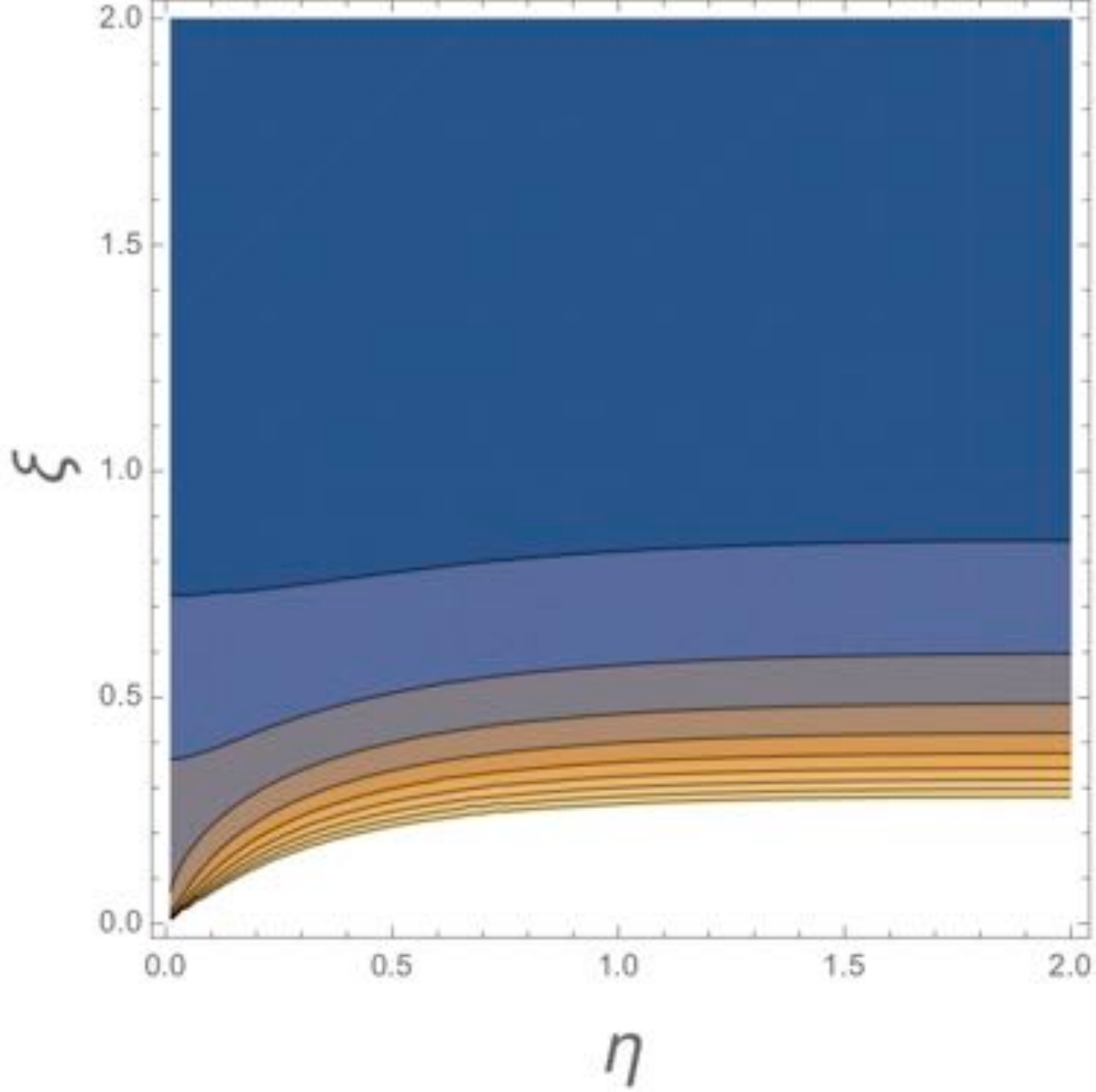}
 \end{minipage} 
 \end{tabular}
\caption{The left figure shows the effective potential $U'$ for the particles with the angular momenta $(L_{\phi_1},L_{\phi_2})=(0,1)$ under the parameter setting $(k_1,k_2,k_3,l_1)=(0,10,-50,1)$. 
The right figure shows the contours of $U'$ for the particles with the same angular momenta. 
The potential does not have a local minimum for the absence of evanescent ergosurface on the $z$ axis. }
\label{fig:L31_I-_L2=infty}
\end{figure}

Moreover, to consider the zero energy limit for particles with the angular momenta $L_{\phi_1}=0$, let us see Fig.~\ref{fig:L31_I-_L2=infty}.
The left figure shows the effective potential $U'$ for the particles with $(L_{\phi_1},L_{\phi_2})=(0,1)$ under the parameter setting $(k_1,k_2,k_3,l_1)=(0,10,-50,1)$. 
The right figure shows the contours of $U'$ for the particles with the same angular momenta. 
The potential $U'$ does not have a local minimum for the absence of evanescent ergosurface on the $z$ axis.
 Massless particles with zero energy are not stably trapped.

\subsubsection{$I_1$}

Let us discuss the existence of stable bound orbits at the interval $I_1$ on the $z$ axis. 
When we expand the potential $U$ at $z=z_2$ as
\begin{eqnarray}
U\simeq \frac{U^{(1)}_{2-}}{z-z_2}+{\cal O}(1).
\end{eqnarray}
The signature of $U^{(1)}_{2-}$ does not depend on angular momenta $l_{\phi_1}$ and $l_{\phi_2}$ of particles, and $U^{(1)}_{2-}<0$ is satisfied within the whole parameter region. 
Therefore $U$ diverges to $\infty$ at $z=z_2$ on $I_1$.
On the other hand, near the horizon $z=0$, $U$ behaves as
\begin{eqnarray}
U\simeq -\frac{l_1^2}{z^2}-\frac{2l_1[(k_3^2+z_3)z_2+k_2^2z_3]}{z_2z_3z}+U^{(0)}_{1+}+U^{(1)}_{1+}z.
\end{eqnarray}
The negativity of the first and second terms makes $U$ diverge to $-\infty$ at the horizon $z=z_1(=0)$, which 
is a pure effect of gravity. 
In order that $U$ has a local minimum near the horizon, $U$ must have a local maximum because $U$ diverges to $\infty$ at $z=z_2$. 
Hence, $U^{(1)}_{1+}$ must be negative, which can be realized 
for sufficiently large $l_{\phi_1}=|-3l_{\phi_2}/2|$ because for $|l_{\phi_2}|\to\infty$,
\begin{eqnarray}
U^{(1)}_{1+}\simeq -\frac{(3k_3^2+l_1)z_2+(3k_2^2+l_1)z_3+3z_2z_3}{16l_1z_2z_3}l_{\phi_2}^2<0.
\end{eqnarray}
Figure~\ref{fig:L31_I2_L2=1000} shows the typical shape of the effective potential $U$.
The left figure shows the effective potential on the $\eta$ axis ($\eta_1\le \eta\le \eta_2$) [on the $z$ axis ($0\le z\le z_2)$] for the particles with the angular momenta 
$(l_{\phi_1},l_{\phi_2})=(-1500,1000)$ under the parameter setting $(k_1,k_2,k_3,l_1)=(0,10,-50,1)$ and 
the right figure shows the contours of $U$ for the particles with the same angular momenta. 
One can see from these figures that $U$ has a negative local minimum on $I_1$ and inside the red curve $U=0$, both massive and massless particles are stably bounded (in the finite region $U\le -m^2/E^2$ and $U\le 0$, respectively). 

 \begin{figure}[h]
 \begin{tabular}{cc}
 \begin{minipage}[t]{0.5\hsize}
\includegraphics[width=7cm,height=7cm]{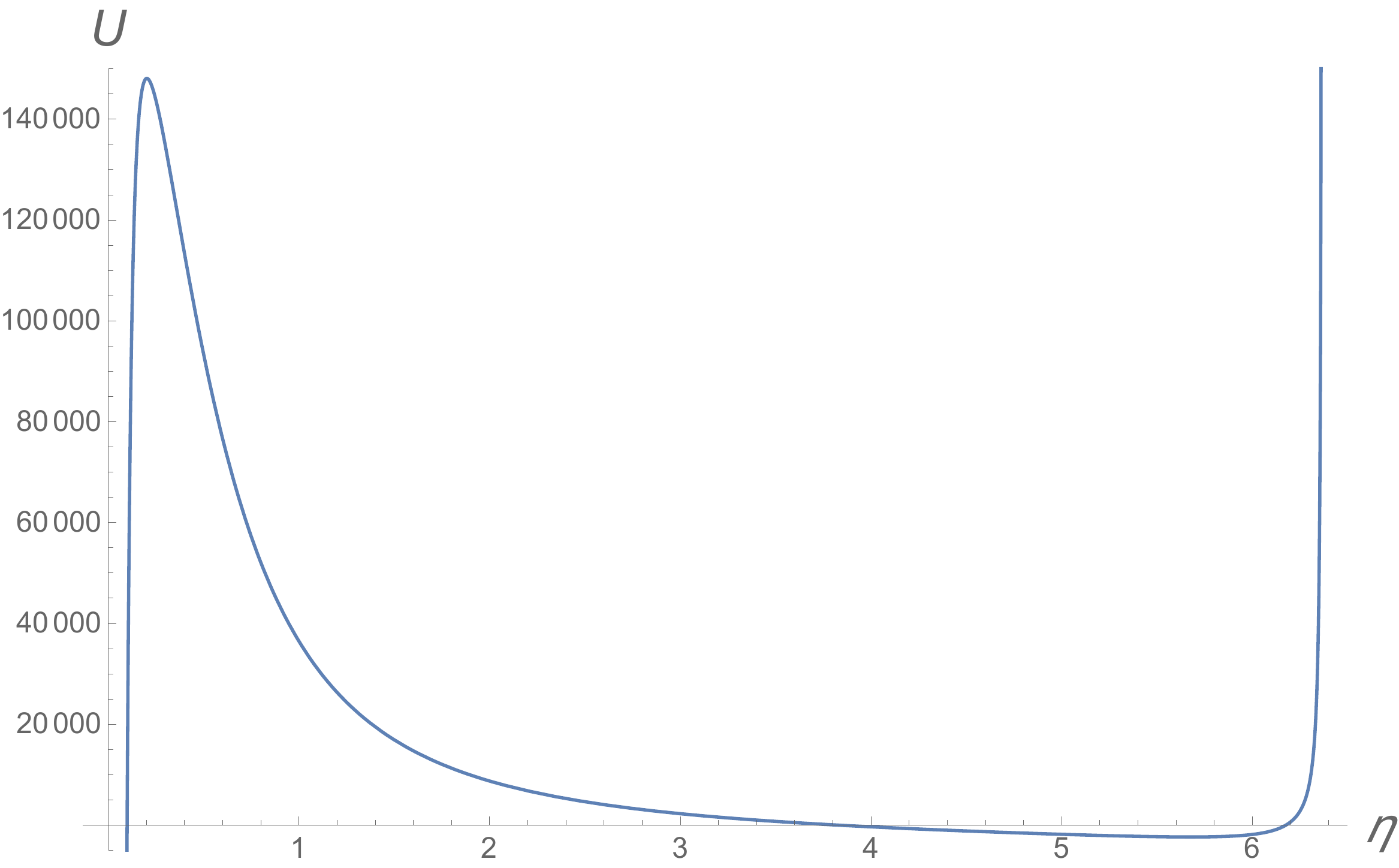}
 \end{minipage} & 
 
 \begin{minipage}[t]{0.5\hsize}
\includegraphics[width=7cm,height=7cm]{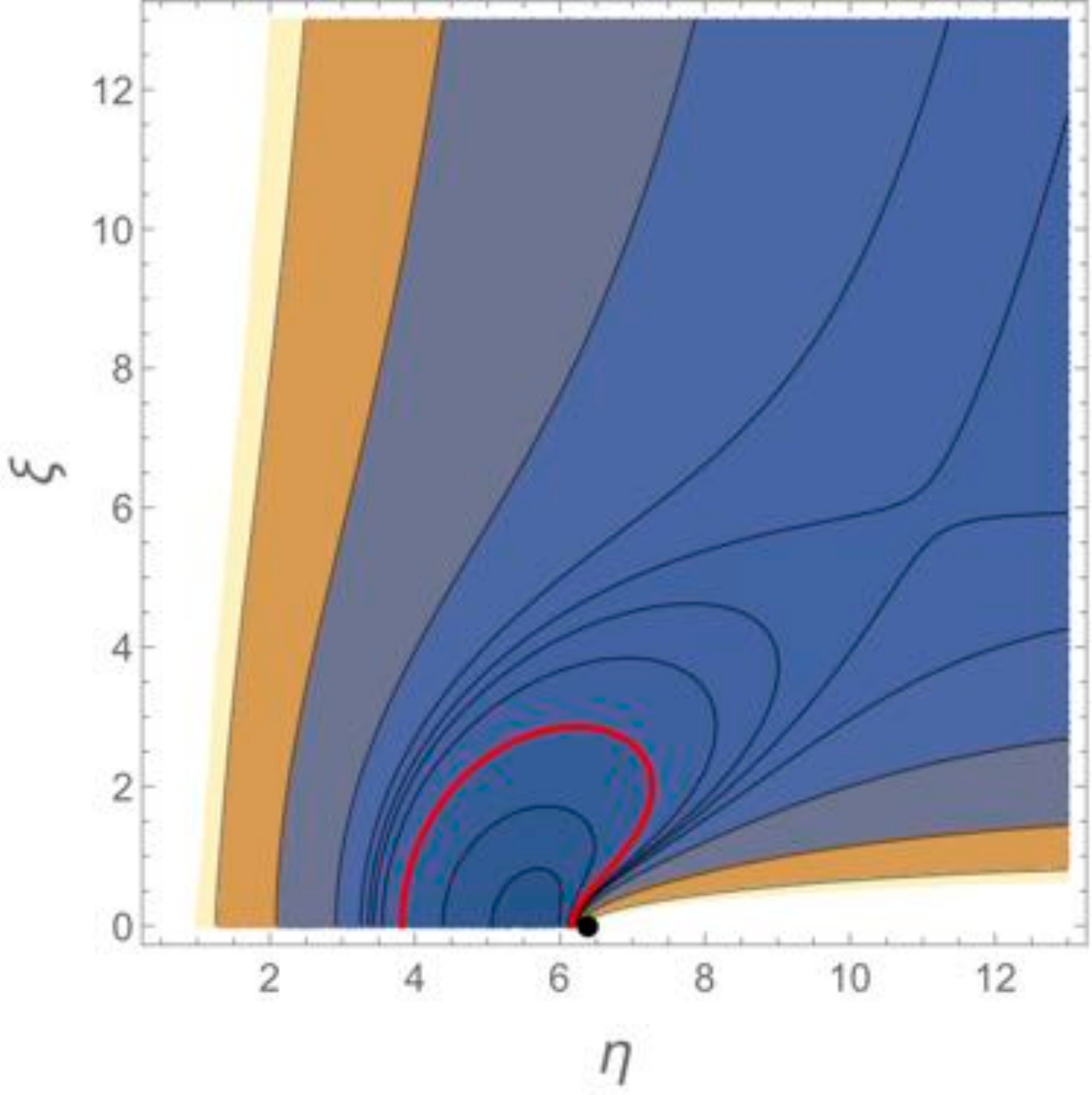}
 \end{minipage}  
 \end{tabular}
\caption{The left figure shows the effective potential $U$ for the particles with the angular momenta $(l_{\phi_1},l_{\phi_2})=(-1500,1000)$ under the parameter setting $(k_1,k_2,k_3,l_1)=(0,10,-50,1)$. 
The right figure shows the contours of $U$ for the particles with the same angular momenta. 
}
\label{fig:L31_I2_L2=1000}
\end{figure}

\begin{figure}[h]
 \begin{tabular}{cc}
 \begin{minipage}[t]{0.5\hsize}
\includegraphics[width=6cm]{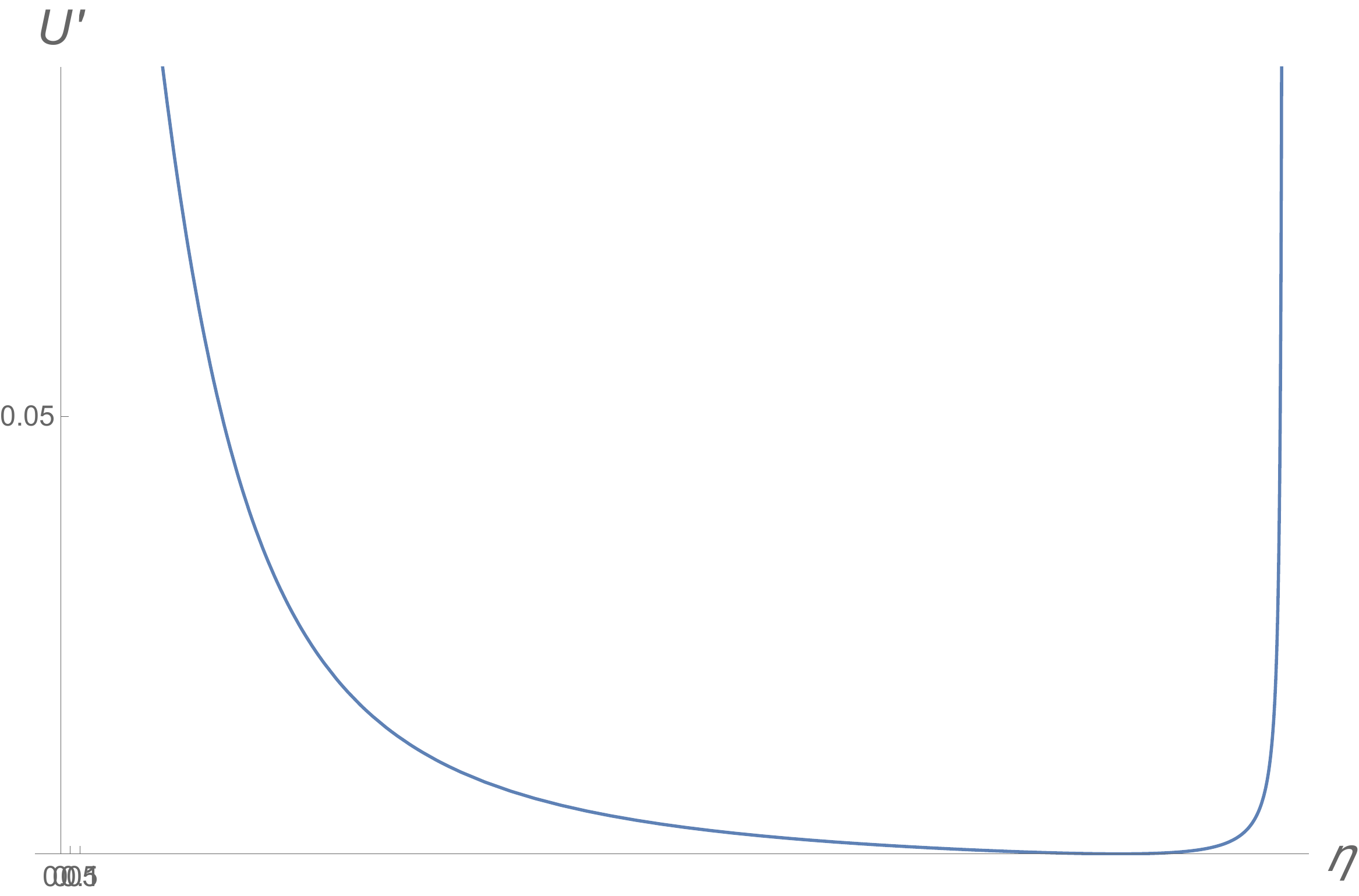}
 \end{minipage} & 
 
 \begin{minipage}[t]{0.5\hsize}
\includegraphics[width=6cm]{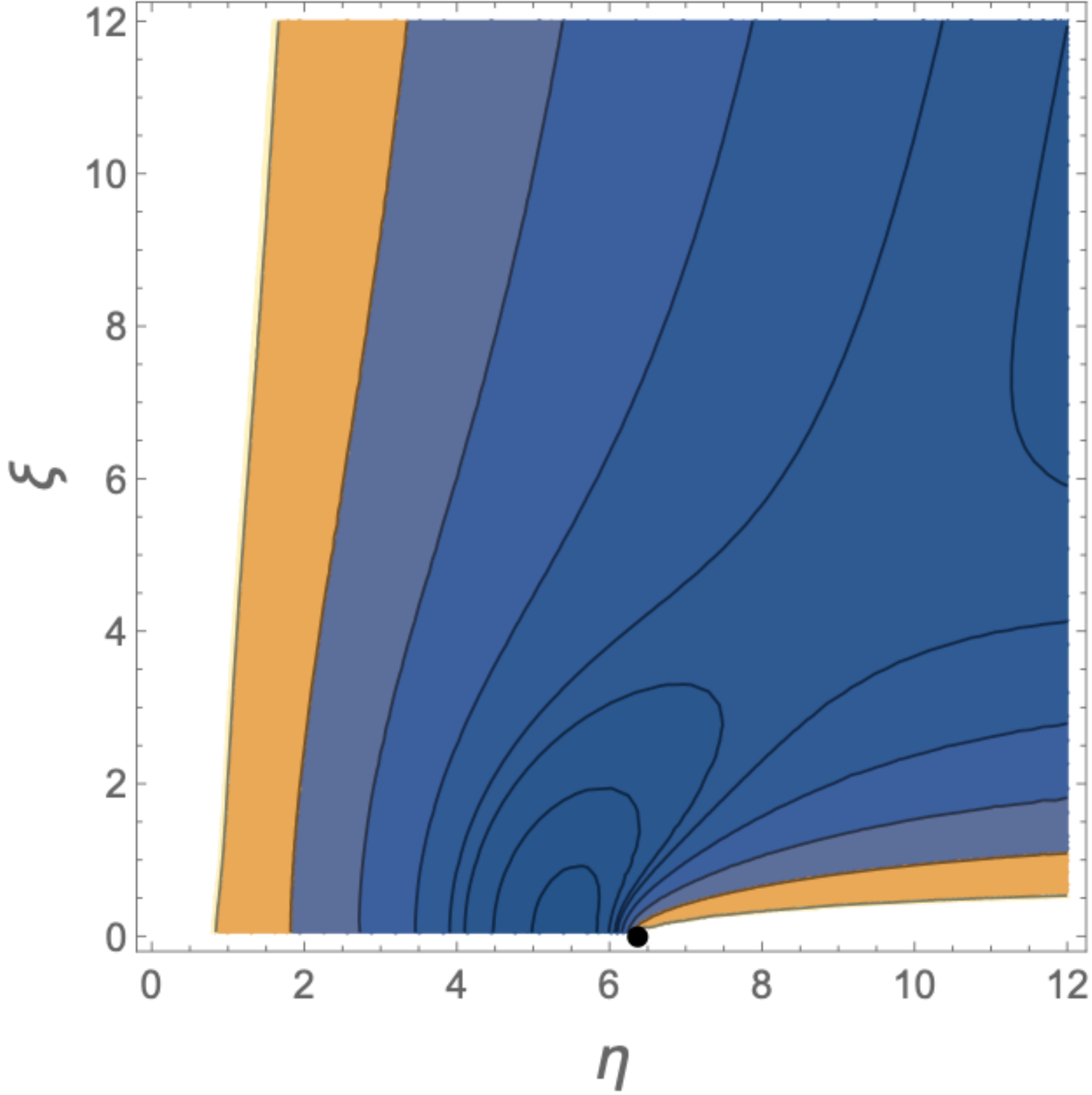}
 \end{minipage}  
 \end{tabular}
\caption{
The left figure shows the typical effective potential $U'$ for $(k_1,k_2,k_3,l_1)=(0,10,-50,1)$ and $(l_{\phi_1},l_{\phi_2})=(3,-2)$.  
The right figure shows the contours of $U'$ for the particles with the same angular momenta. The potential has a local minimum whose values are zero on the evanescent ergosurface. 
 }

\label{fig:L31_I1_L2=infty}
\end{figure}

To consider the geodesic motion of the massless particles with zero-energy limit $E\to 0$, let us see Fig.~\ref{fig:L31_I1_L2=infty}.
The left figure shows the typical effective potential $U'$ for the particles with the angular momenta $L_{\phi_1}/L_{\phi_2}=-3/2$ under the parameter setting $(k_1,k_2,k_3,l_1)=(0,10,-50,1)$ and $(L_{\phi_1},L_{\phi_2})=(3,-2)$. 
The right figure shows the contours of $U'$ for the particles with the same angular momenta. 
The potential has a zero local minimum on the evanescent ergosurface, where massless particles with zero energy are stably trapped.

\subsubsection{$I_2$}
Finally, let us discuss the existence of stable bound orbits on $I_2$. 
This type of an interval does not exist for the black lens with $L(2,1)$ topology. 
When we expand the potential $U$ at $z=z_2,\ z_3$ as
\begin{eqnarray*}
U\simeq \frac{U^{(1)}_{i}}{z-z_i},\ (i=2+,3-)
\end{eqnarray*}
Regardless of the angular momenta $l_{\phi_1}$ and $l_{\phi_2}$ of particles, $U^{(1)}_{2-}$ and $U^{(1)}_{3+}$ are positive and negative, respectively, within the parameter region ${\cal D}$. 
Therefore $U$ diverges to $\infty$ at $z=z_2$ and $z=z_3$ on $I_2$. 
As is already expected, $U$ has a local minimum on $I_2$. 

Figure~\ref{fig:L31_I2_L2=1000} shows the typical shape of the effective potential $U$.
The left figure shows the effective potential on the $\eta$ axis ($\eta_2\le \eta\le \eta_3$) [on the $z$ axis ($z_2\le z\le z_3)$] for the particles with the angular momenta 
$(l_{\phi_1},l_{\phi_2})=(-2000,1000)$ under the parameter setting $(k_1,k_2,k_3,l_1)=(0,10,-50,1)$ and 
the right figure shows the contours of $U$ for the particles with the same angular momenta. 
One can see from these figures that $U$ has a negative local minimum on $I_2$ and inside the red curve $U=0$, both massive and massless particles are stably bounded (in the finite region $U\le -m^2/E^2$ and $U\le 0$, respectively).

 \begin{figure}[h]
 \begin{tabular}{cc}
 \begin{minipage}[t]{0.5\hsize}
\includegraphics[width=6cm]{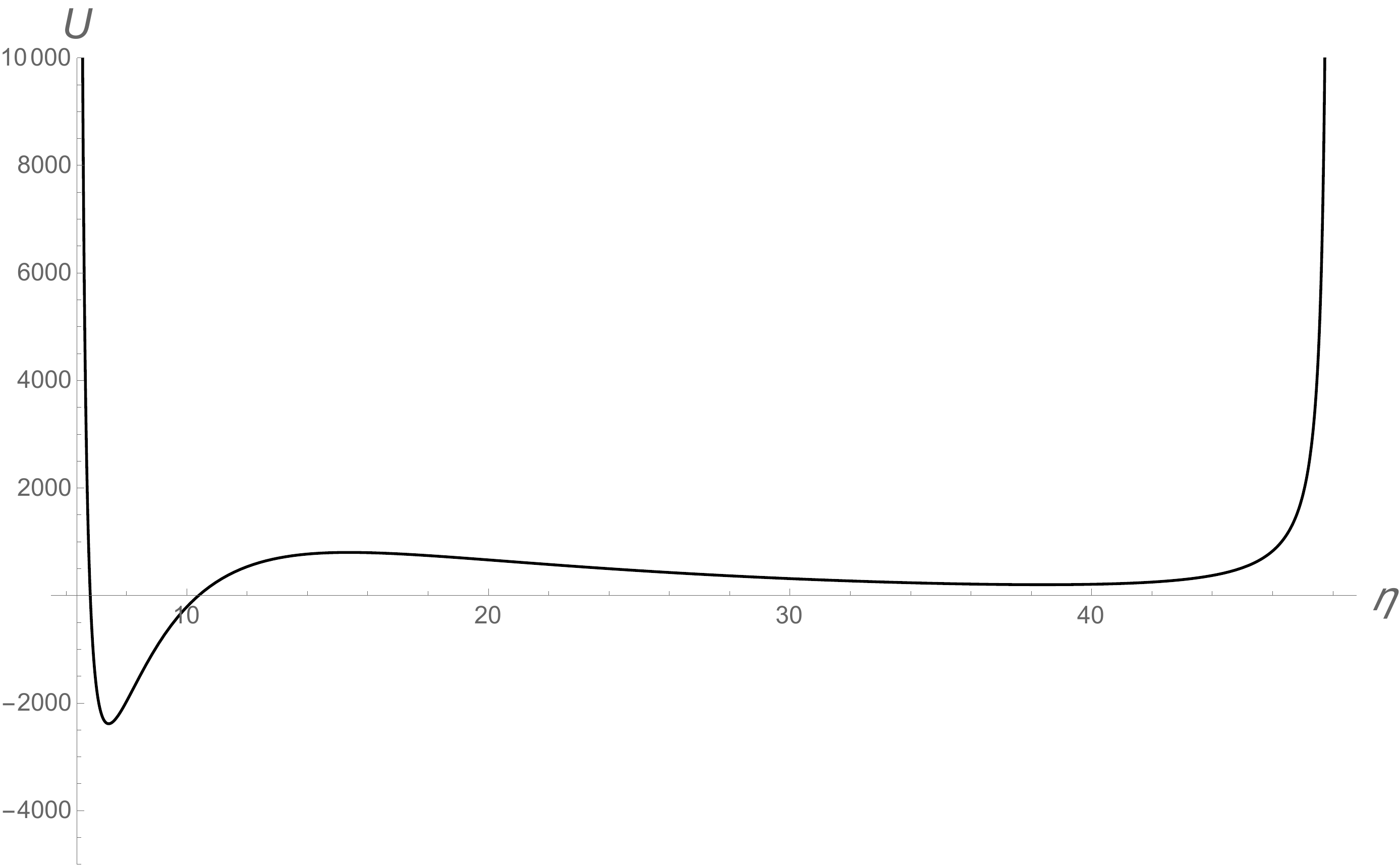}
 \end{minipage} & 
 
 \begin{minipage}[t]{0.5\hsize}
\includegraphics[width=6cm]{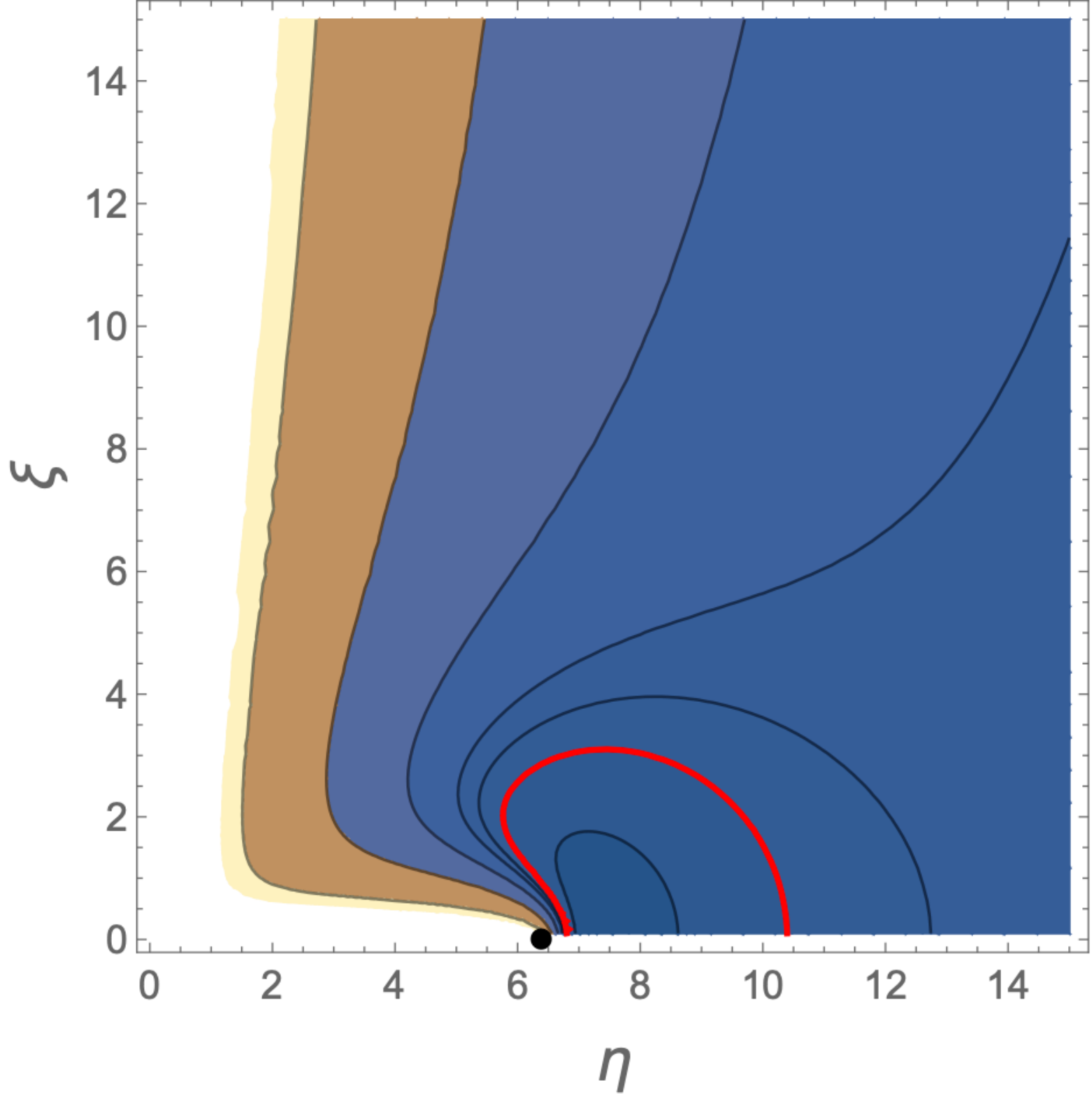}
 \end{minipage}  
 \end{tabular}
\caption{The left figure shows the effective potential $U$ for the particles with the angular momenta $(l_{\phi_1},l_{\phi_2})=(-2000,1000)$ under the parameter setting $(k_1,k_2,k_3,l_1)=(0,10,-50,1)$. 
The right figure shows the contours of $U$ for the particles with the same angular momenta.  }
\label{fig:L31_I2_L2=1000}
\end{figure}

\begin{figure}[h]
 \begin{tabular}{cc}
 \begin{minipage}[t]{0.5\hsize}
\includegraphics[width=7cm,height=7cm]{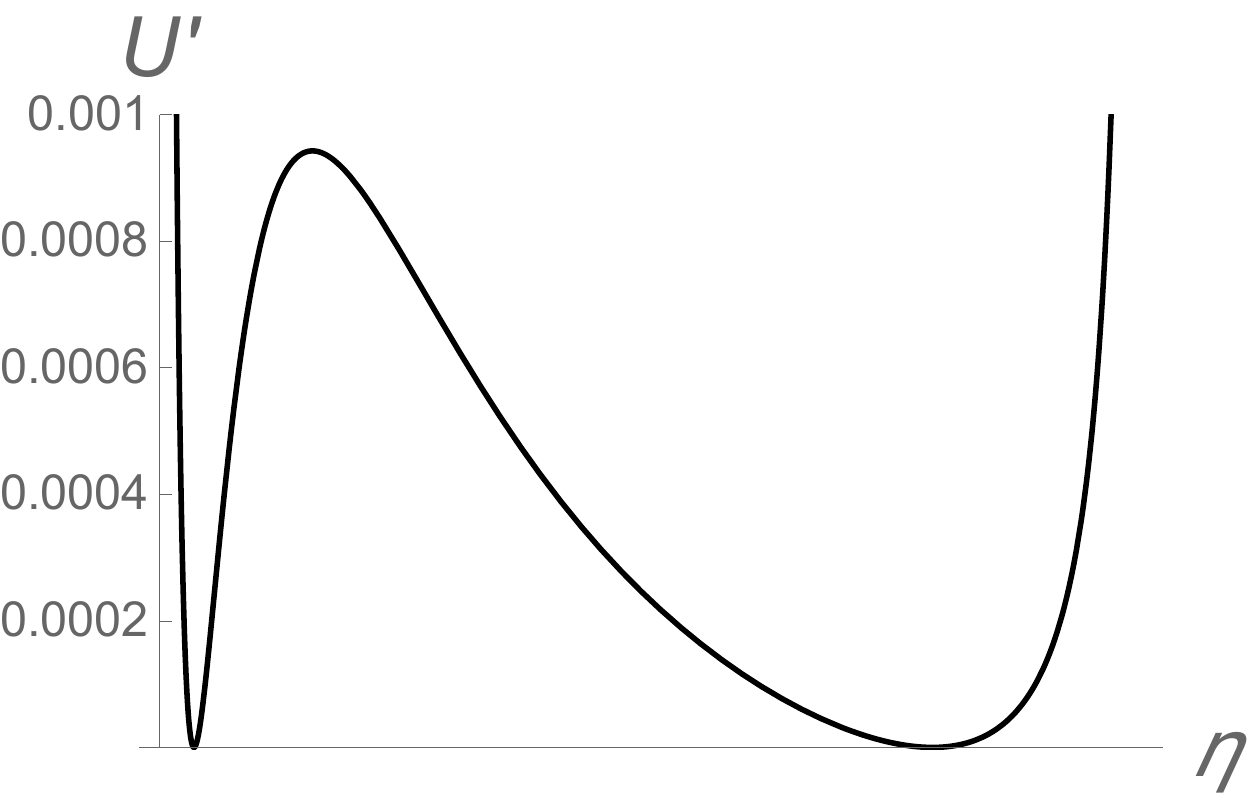}
 \end{minipage} & 
 
 \begin{minipage}[t]{0.5\hsize}
\includegraphics[width=7cm,height=7cm]{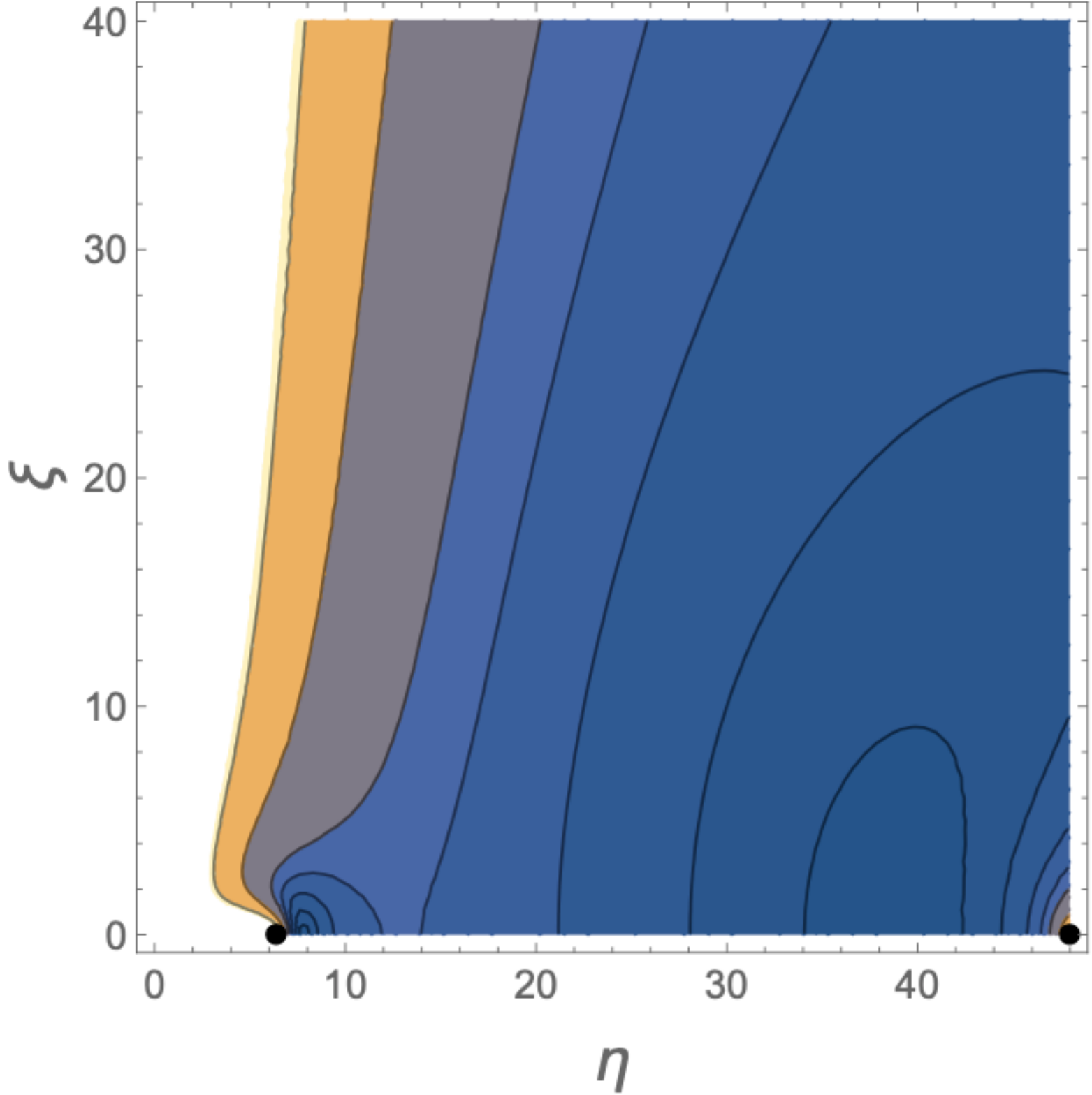}
 \end{minipage}  
 \end{tabular}
\caption{The left figure shows the effective potential $U'$ for the particles with the angular momenta $(L_{\phi_1},L_{\phi_2})=(-2,1)$ under the parameter setting $(k_1,k_2,k_3,l_1)=(0,10,-50,1)$. 
The right figure shows the contours of $U'$ for the particles with the same angular momenta.  The potential has local minima whose values are zero at the evanescent ergosurfaces. }

\label{fig:L31_I2_L2=infty}
\end{figure}

Moreover, to consider the zero energy limit for particles with the angular momenta $L_{\phi_1}/L_{\phi_2}=-2$, let us see Fig.~\ref{fig:L31_I2_L2=infty}.
The left figure shows the effective potential $U'$ for the particles with $(L_{\phi_1},L_{\phi_2})=(-2,1)$ under the parameter setting $(k_1,k_2,k_3,l_1)=(0,10,-50,1)$. 
The right figure shows the contours of $U'$ for the particles with the same angular momenta. 
The potential has two local minima, whose values are zero at the evanescent ergosurfaces.

%%%%%%%%%%%%%%%%%%%%%%%%%%%%%%%%%%%%%%%%%%%%
%%%%%%%%%%%%%%%%%%%%%%%%%%%%%%%%%%%%%%%%%%%%
%%%%%%%%%%%%%%%%%%%%%%%%%%%%%%%%%%%%%%%%%%%%

%                                << summary>>

%%%%%%%%%%%%%%%%%%%%%%%%%%%%%%%%%%%%%%%%%%%%
%%%%%%%%%%%%%%%%%%%%%%%%%%%%%%%%%%%%%%%%%%%%
%%%%%%%%%%%%%%%%%%%%%%%%%%%%%%%%%%%%%%%%%%%%
\section{Summary and Discussions}\label{sec:summary}

In this paper, we have analyzed the geodesics for the supersymmetric black lenses with $L(2,1)$ and $L(3,1)$ topologies in the five-dimensional minimal supergravity. 
 We have reduced the geodesic motion of particles to a two-dimensional potential problem, and from the two-dimensional contour plots of the potential, we have clarified the existence of the stable bound orbits for the massive and massless particles. 
In the previous work~\cite{Tomizawa:2019egx} on the black lens with $L(2,1)$ topology, where one-dimensional potential was used, we found the stable bound orbits of particles on the $z$ axis corresponding to two rotational axes ($\eta$ axis and $\xi$ axis). 
In this work, we have found that there also exist stable bound orbits in a place away from the $z$ axis as well as on the axis. 
Moreover, we have discussed the range such that stable bound orbits exist on the $z$ axis and have shown that there exist no stable bound orbits in the asymptotic region $z\to \pm\infty$, at least, on the $z$ axis. 
This result suggests that there exists the outermost stable circular orbit as the boundary of stable circular orbits.

\medskip
In particular, we have also discussed the geodesic motion for massless particles with zero energy. 
We have proved that for such particles, the potential always has a local minimum at the intersection of the evanescent ergosurface and the curve $G(\eta,\xi,L_{\phi_1},L_{\phi_2})=0$ in the $(\eta,\xi)$ plane, where, in particular, for particles with the angular momenta satisfying either condition $L_{\phi_2}=0$ or $J=0$, the intersection exists at least, at the rotational axes. 
This means that such particles are stably trapped on the geodesics on the the evanescent ergosurface, which is exactly consistent with the mathematical result~\cite{Eperon:2016cdd}, 
which proved that on an evanescent ergosurface, massless particles with zero energy  move along stable trapped null geodesics. 
As shown in Ref.~\cite{Eperon:2016cdd}, for the microstate geometry, the existence of evanescent ergosurfaces leads to some nonlinear instability. 
This result does not apply to the solution with a black hole horizon but the presence of stable bound orbits of particles with zero energy even  in a black lens background may exhibit corresponding nonlinear instability.

 %%%%%%%%%%%%%%%%%%%%%%%%%%%%%%%%%%%%%%%%%%%%
%%%%%%%%%%%%%%%%%%%%%%%%%%%%%%%%%%%%%%%%%%%%
%%%%%%%%%%%%%%%%%%%%%%%%%%%%%%%%%%%%%%%%%%%%

%                                << acknowledgments>>

%%%%%%%%%%%%%%%%%%%%%%%%%%%%%%%%%%%%%%%%%%%%
%%%%%%%%%%%%%%%%%%%%%%%%%%%%%%%%%%%%%%%%%%%%
%%%%%%%%%%%%%%%%%%%%%%%%%%%%%%%%%%%%%%%%%%%%

\acknowledgments
This work was supported by the Grant-in-Aid for Scientific Research (C) [JSPS KAKENHI Grant Number~17K05452~(S.T.)] and the Grant-in-Aid for Early-Career Scientists [JSPS KAKENHI Grant Number~JP19K14715~(T.I.)] from the Japan Society for the Promotion of Science. S.T. is also supported from Toyota Institute of Technology Fund for
Research Promotion (A).

%%%%%%%%%%%%%%%%%%%%%%%%%%%%%%%%%%%%%%%%%%%%
%%%%%%%%%%%%%%%%%%%%%%%%%%%%%%%%%%%%%%%%%%%%
%%%%%%%%%%%%%%%%%%%%%%%%%%%%%%%%%%%%%%%%%%%%

%                                << Reference>>

%%%%%%%%%%%%%%%%%%%%%%%%%%%%%%%%%%%%%%%%%%%%
%%%%%%%%%%%%%%%%%%%%%%%%%%%%%%%%%%%%%%%%%%%%
%%%%%%%%%%%%%%%%%%%%%%%%%%%%%%%%%%%%%%%%%%%%

\end{document}